%% file: ADGS.tex
\documentclass[11pt]{article}
\pdfoutput=1
\usepackage[utf8]{inputenc}
\usepackage{graphicx} 
\usepackage{epstopdf} 
\usepackage{epsfig,cite,color,amsmath,amssymb,amsfonts,url}
\usepackage[title]{appendix}

\pdfoutput=1
\allowdisplaybreaks

\newcommand{\beq}{\begin{eqnarray}}
\newcommand{\eeq}{\end{eqnarray}}
\newcommand{\gsim}{\raisebox{-0.13cm}{~\shortstack{$>$ \\[-0.07cm]
      $\sim$}}~}

\def\nn{\nonumber}
\def\be{\begin{equation}}
\def\ee{\end{equation}}
\newcommand{\bea}{\begin{eqnarray}}
\newcommand{\eea}{\end{eqnarray}}
\newcommand{\bdm}{\begin{displaymath}}
\newcommand{\edm}{\end{displaymath}}
\long\def\symbolfootnote[#1]#2{\begingroup%
\def\thefootnote{\fnsymbol{footnote}}\footnote[#1]{#2}\endgroup}

\def\sq2{\sqrt{2}}
\def\drbar{\overline{\rm DR}}

\def\smallsm{\scriptscriptstyle{SM}}

\def\smallS{{\scriptscriptstyle S}}

\def\tb{\tan\beta}
\def\gl{\tilde{g}}
\def\mg{m_{\gl}}
\def\g{\mg^2}

\def\msusy{M}
\def\msquark{M_S}
\def\invm{M^2_{\phi\chi}}
\def\invhh{M_{hh}}
\def\szhat{\invm}
\def\szhathh{\invhh^2}

\newcommand{\smallz}{{\scriptscriptstyle Z}} 
\newcommand{\smallw}{{\scriptscriptstyle W}} %
\newcommand{\smallH}{{\scriptscriptstyle H}} %
\newcommand{\smallr}{{\scriptscriptstyle R}} %
\newcommand{\smalll}{{\scriptscriptstyle L}} %
\newcommand{\smalla}{{\scriptscriptstyle A}}
\newcommand{\mz}{m_\smallz}
\newcommand{\mw}{m_\smallw}

\newcommand{\ma}{m_\smalla}
\newcommand{\muF}{\mu_{\scriptscriptstyle F}}
\newcommand{\muR}{\mu_\smallr}
\newcommand{\cupc}{C_{\Delta}^{h\phi\chi} }
\newcommand{\cdpc}{C_{\Delta}^{H\phi\chi} }
\newcommand{\LET}{{\scriptscriptstyle{\rm LET}}}

\def\as{\alpha_s}

\def\mt{m_t}

\def\tu{m_{\tilde{t}_1}^2}
\def\td{m_{\tilde{t}_2}^2}
\def\tul{m_{\tilde{t}_1}}

\def\ti{m_{\tilde{t}_i}^2}
\def\tuq{m_{\tilde{t}_1}^4}
\def\tdq{m_{\tilde{t}_2}^4}
\def\tuc{m_{\tilde{t}_1}^6}
\def\tdc{m_{\tilde{t}_2}^6}
\def\sdt{s_{2\theta_t}}
\def\cdt{c_{2\theta_t}}
\def\cdtt{c_{2\theta_t}}

\def\dtl{d^t_\smalll}
\def\dtr{d^t_\smallr}
\def\dtuu{d^t_{11}}
\def\dtud{d^t_{12}}
\def\dtdd{d^t_{22}}

\def\xl{x_\smalll}
\def\dbl{d^b_\smalll}
\def\dbr{d^b_\smallr}
\def\mb{m_b}
\def\sbl{{\tilde b_\smalll}}
\def\sbr{{\tilde b_\smallr}}
\def\bl{m_\sbl^2}
\def\blq{m_\sbl^4}
\def\bls{m_\sbl^6}
\def\br{m_\sbr^2}
\def\brq{m_\sbr^4}
\def\brs{m_\sbr^6}

\def\blob{Z}
\def\DVtu{\frac{\partial \blob}{\partial \tu}}
\def\DVtd{\frac{\partial \blob}{\partial \td}}
\def\DVcdtq{\frac{\partial \blob}{\partial \cdtt^2}}
\def\DVt{\frac{\partial \blob}{\partial \mt^2 }}
\def\DVtt{\frac{\partial^{\,2} \blob}{(\partial \mt^2)^2 }}
\def\DVttu{\frac{\partial^{\,2} \blob}{\partial \mt^2 \partial \tu }}
\def\DVttd{\frac{\partial^{\,2} \blob}{\partial \mt^2 \partial \td }}
\def\DVtutu{\frac{\partial^{\,2} \blob}{(\partial \tu)^2 }}
\def\DVtutd{\frac{\partial^{\,2} \blob}{\partial \tu \partial \td }}
\def\DVtdtd{\frac{\partial^{\,2} \blob}{(\partial \td)^2}}
\def\DVcdtqtu{\frac{\partial^{\,2} \blob}{\partial \cdtt^2 \partial \tu }}
\def\DVcdtqtd{\frac{\partial^{\,2} \blob}{\partial \cdtt^2 \partial \td }}
\def\DVcdtqt{\frac{\partial^{\,2} \blob}{\partial \cdtt^2 \partial \mt^2 }}
\def\DVcdtqcdtq{\frac{\partial^{\,2} \blob}{(\partial \cdtt^2)^2 }}

\newcommand{\Eqref}[1]{eq.\,(\ref{#1})}

\renewcommand{\bar}{\overline}

\newcommand{\f}[2]{\frac{#1}{#2}}

\begin{document}
\begin{titlepage}


{\flushright{
        \begin{minipage}{3cm}
          RM3-TH/16-1
        \end{minipage}        }

}
\renewcommand{\thefootnote}{\fnsymbol{footnote}}
\vskip 2cm
\begin{center}
\boldmath
{\LARGE\bf NLO-QCD corrections to Higgs pair production}
\vskip 0.3cm
{\LARGE\bf in the MSSM}\unboldmath
\vskip 1.cm
{\Large{A.~Agostini$^{a,b}$, G.~Degrassi$^{a,b}$, 
R.~Gr\"ober$^{b}$ and P.~Slavich$^{c,d}$}}
\vspace*{8mm} \\
{\sl ${}^a$
    Dipartimento di Matematica e Fisica, Universit\`a di Roma Tre \\
    Via della Vasca Navale~84, I-00146 Rome, Italy}
\vspace*{2mm}\\
{\sl ${}^b$ INFN, Sezione di Roma Tre, Via della Vasca Navale~84, I-00146 Rome, Italy}
\vspace*{2mm}\\
{\sl ${}^c$ LPTHE, UPMC Univ.~Paris 06,
  Sorbonne Universit\'es, 4 Place Jussieu, F-75252 Paris, France}
\vspace*{2mm}\\
{\sl ${}^d$ LPTHE, CNRS, 4 Place Jussieu, F-75252 Paris, France }
\end{center}
\symbolfootnote[0]{{\tt e-mail:}}
\symbolfootnote[0]{{\tt agostini@fis.uniroma3.it}}
\symbolfootnote[0]{{\tt degrassi@fis.uniroma3.it}}
\symbolfootnote[0]{{\tt groeber@roma3.infn.it}}
\symbolfootnote[0]{{\tt slavich@lpthe.jussieu.fr}}

\vskip 0.7cm

\begin{abstract}
We take a step towards a complete NLO-QCD determination of the
production of a pair of Higgs scalars in the MSSM. Exploiting a
low-energy theorem that connects the Higgs-gluon interactions to the
derivatives of the gluon self-energy, we obtain analytic results for
the one- and two-loop squark contributions to Higgs pair production in
the limit of vanishing external momenta.  We find that the two-loop
squark contributions can have non-negligible effects in MSSM scenarios
with stop masses below the TeV scale. We also show how our results can
be adapted to the case of Higgs pair production in the NMSSM.
\end{abstract}
\vfill
\end{titlepage}    
\setcounter{footnote}{0}


\section{Introduction}
\label{sec:intro}

After the discovery of a Higgs boson in Run 1 of the
LHC~\cite{Aad:2012tfa,Chatrchyan:2012ufa}, one of the major goals of
Run 2 is the experimental exploration of its properties. In Run 1, the
couplings of the Higgs boson to fermions and to gauge bosons have already
been measured, and found to be compatible with the predictions of the
Standard Model (SM) within an experimental accuracy of (10\! --
\!20)\%~\cite{ATLAS-CONF-2015-044}.  On the other hand, the
self-couplings of the Higgs boson, which are accessible in multi-Higgs
production processes, have not been probed yet. While a measurement of
the quartic Higgs self-coupling lies beyond the reach of the
LHC~\cite{Plehn:2005nk, Binoth:2006ym}, previous studies showed that
the Higgs pair production process, and hence the trilinear Higgs
self-coupling, might be accessible for high integrated luminosities in
the $b\bar{b} \gamma \gamma$~\cite{Baur:2003gp, Baglio:2012np,
  Yao:2013ika, Barger:2013jfa, Azatov:2015oxa, Lu:2015jza}, $b\bar{b}
\tau \bar{\tau}$~\cite{Dolan:2012rv, Baglio:2012np}, $b\bar{b}W^+
W^-$~\cite{Papaefstathiou:2012qe} and
$b\bar{b}b\bar{b}$~\cite{deLima:2014dta, Wardrope:2014kya, Behr:2015oqq} final
states.

Not only is Higgs pair production interesting as a probe of the
trilinear Higgs self-coupling in the SM, but it also can help
constrain the SM extensions. First limits on scenarios with strongly
increased cross section, which occurs, e.g., in models with novel $hh
t\bar{t}$ coupling \cite{Dib:2005re, Grober:2010yv, Contino:2012xk},
or if the Higgs boson pair is produced through the decay of a heavy
new resonance, have been given in refs.~\cite{CMS:2014ipa,
  Aad:2014yja, Khachatryan:2015yea, Aad:2015uka, Aad:2015xja}.

In the minimal supersymmetric extension of the SM (MSSM) the Higgs
sector consists of two SU(2) doublets, $H_1$ and $H_2$, whose relative
contribution to electroweak (EW) symmetry breaking is determined by
the ratio of vacuum expectation values (VEVs) of their neutral
components, $\tan \beta\equiv v_2/v_1$. The spectrum of physical Higgs
bosons is richer than in the SM, consisting of two neutral scalars,
$h$ and $H$, one neutral pseudoscalar, $A$, and two charged scalars,
$H^{\pm}$.  The couplings of the scalars to matter fermions and gauge
bosons, as well as their self-couplings, differ in general from the SM
ones. However, in the so-called decoupling limit of the MSSM Higgs
sector, $\ma \gg \mz$, the lightest scalar $h$ has SM-like couplings
and can be identified with the particle discovered at the LHC, with
$m_h \approx 125$~GeV~\cite{Aad:2015zhl}.

The dominant mechanism for Higgs pair production in the MSSM is gluon
fusion,\footnote{For the single production of neutral Higgs bosons
  with enhanced couplings to down-type fermions, the $b\bar{b}$
  annihilation process dominates over gluon fusion for intermediate to
  large values of $\tb$. In contrast, for Higgs pair production this
  is only the case in very limited regions of the MSSM parameter
  space~\cite{Cao:2013si, Han:2013sga}.} mediated by loops involving
the top and bottom quarks and their superpartners, the stop and
sbottom squarks.  Only for relatively light squarks, with masses below
the TeV scale, do the squark contributions lead to sizeable effects on
the cross section for the production of SM-like Higgs
pairs~\cite{Batell:2015koa}. Direct searches leave several corners of
parameter space for light stops open, e.g.~for reduced branching
ratios or difficult kinematic configurations~\cite{Grober:2014aha,
  Aad:2015pfx, Belyaev:2015gna}.  However, the measured value of $m_h$
implies either stop masses in the multi-TeV range or a large and
somewhat tuned left-right mixing in the stop mass matrix. Scenarios
allowing for light stops are thus restricted to the latter
possibility.

Due to the extended Higgs spectrum of the MSSM, a pair of light
scalars can also be produced resonantly through the $s$-channel
exchange of a heavy scalar, leading to a sizeable increase in the
cross section~\cite{Plehn:1996wb,Dawson:1998py,Djouadi:1999rca,
  Muhlleitner:2000jj}. In addition, mixed scalar/pseudoscalar pairs,
and pairs of pseudoscalars, can as well be produced in gluon
fusion. In this paper, however, we will restrict our attention to the
production of scalar pairs.
 

In the SM, the leading-order (LO) cross section for Higgs pair
production via gluon fusion, fully known since the late
eighties~\cite{Glover:1987nx}, is subject to large radiative
corrections. The next-to-leading order (NLO) QCD contributions of
diagrams involving top quarks were computed in the late nineties in
the limit of infinite top mass $m_t$, or, equivalently, of vanishing
external momenta~\cite{Dawson:1998py}.  Whereas this approximation was
shown to work quite well for single Higgs
production~\cite{Spira:1995rr}, it can be expected to be less
effective for pair production, due to the larger energy scale that
characterizes the latter process. Unfortunately, an exact two-loop
calculation of the ``box'' form factor that contributes to Higgs pair
production at the NLO in QCD is currently not available. In contrast,
the ``triangle'' form factor entering diagrams where a single
($s$-channel) Higgs boson splits into a Higgs pair can be borrowed
from the the calculation of single Higgs
production~\cite{Spira:1995rr, Harlander:2005rq, Anastasiou:2006hc,
  Aglietti:2006tp}.

In order to improve the NLO result for Higgs pair production in the
SM, ref.~\cite{Dawson:1998py} factored out the LO cross section with
the full top-mass dependence. The uncertainties of this approach were
estimated to be of $\mathcal{O}(10 \,\%)$ in
refs.~\cite{Grigo:2013rya, Grigo:2015dia,Frederix:2014hta,
  Maltoni:2014eza}. The next-to-next-to leading order (NNLO)
contributions in the heavy-top limit were computed in
refs.~\cite{deFlorian:2013uza, deFlorian:2013jea, Grigo:2014jma}. Soft
gluon resummation at next-to-next-leading logarithmic (NNLL) order was
performed in refs.~\cite{Shao:2013bz, deFlorian:2015moa}.
Furthermore, NLO contributions in the heavy-top limit have been
computed for the SM extended with dimension six
operators~\cite{Grober:2015cwa}, for an additional scalar
singlet~\cite{Dawson:2015haa} and for the two-Higgs-doublet
model~\cite{Hespel:2014sla}.

In the case of the MSSM, the triangle form factor~\footnote{In the
  MSSM, loop topologies other than triangle and box contribute to
  scalar pair production, due to the existence of quartic interactions
  involving squarks. With a slight abuse of language, in the following
  we denote as ``triangle'' all diagrams that involve the $s$-channel
  exchange of a single scalar, and as ``box'' all of the remaining
  diagrams.} that contributes to the production of a scalar pair at
the NLO can again be borrowed from the calculation of single-scalar
production. In particular, the contributions of two-loop diagrams
involving only quarks and gluons can be adapted from the corresponding
SM results~\cite{Spira:1995rr, Harlander:2005rq, Anastasiou:2006hc,
  Aglietti:2006tp} via a rescaling of the Higgs-quark couplings. The
contributions of two-loop diagrams involving only squarks and gluons
are fully known~\cite{Anastasiou:2006hc, Aglietti:2006tp,Muhlleitner:2006wx,
  Bonciani:2007ex}. In contrast, an exact calculation of the two-loop
diagrams involving quarks, squarks and gluinos -- which can involve up
to five different masses -- is still missing. Calculations based on a
combination of numerical and analytic methods were presented in
refs.~\cite{Anastasiou:2008rm, Muhlleitner:2010nm}, but neither
explicit formulae nor computer codes implementing the results of those
calculations have been made available so far. Approximate results for
the quark-squark-gluino contributions can however be obtained in the
presence of some hierarchy between the relevant masses. The
top-stop-gluino contributions were computed in the vanishing
Higgs-mass limit (VHML) in refs.~\cite{Harlander:2003bb,
  Harlander:2004tp, Degrassi:2008zj}, and both the top-stop-gluino and
bottom-sbottom-gluino contributions were computed in the limit of
heavy superparticles -- but without assuming a hierarchy between the
Higgs mass and the quark mass -- in refs.~\cite{Degrassi:2010eu,
  Harlander:2010wr, Degrassi:2012vt}. In particular, the calculation
in ref.~\cite{Degrassi:2008zj} relied on a low-energy theorem
(LET)~\cite{Ellis:1975ap, Shifman:1979eb, Kniehl:1995tn}, connecting
the amplitude for Higgs-gluon-gluon interaction to the derivatives of
the gluon self-energy with respect to the Higgs fields, to provide
explicit and compact analytic formulae for the top-stop-gluino
contributions to the triangle form factor in the VHML.

\vfill
\newpage

For what concerns the box form factor, in the MSSM the contributions
of one-loop diagrams involving quarks differ from their SM
counterparts by a rescaling of the Higgs-quark couplings, and their
calculation must be extended to account for the possibility of two
different scalars in the final state~\cite{Plehn:1996wb}. The
contributions of one-loop diagrams involving squarks have been
computed in refs.~\cite{Belyaev:1999mx, BarrientosBendezu:2001di} (see
also ref.~\cite{Batell:2015koa}). Going beyond the LO calculation, the
contributions of two-loop diagrams involving top quarks and gluons in
the heavy-top limit can be adapted from the corresponding SM results
via a rescaling of the Higgs-top couplings~\cite{Dawson:1998py}. On
the other hand, the diagrams involving bottom quarks -- whose effect
is negligible in the SM, but can become relevant in the MSSM where at
least one of the scalars has $\tb$-enhanced couplings to down-type
quarks -- are known only at one loop, because the heavy-quark limit
adopted in the existing NLO calculations cannot, of course, be applied
to them. Finally, no calculation of the contributions to the box form
factor from two-loop diagrams involving squarks has, to our knowledge,
been presented so far.

\bigskip

In this paper we take a step towards a complete NLO-QCD determination
of the production of a pair of Higgs scalars in the MSSM. Relying on
the same LET as in ref.~\cite{Degrassi:2008zj}, we obtain analytic
results for the contributions to the box form factor from one- and
two-loop diagrams involving top quarks and stop squarks in the limit
of vanishing external momenta. We also obtain, by direct calculation
of the relevant two-loop diagrams, the subset of bottom/sbottom
contributions that involve the $D$-term-induced EW Higgs-squark
coupling and survive in the limit of vanishing bottom mass.  To assess
the importance of the newly-computed corrections, we include the
squark contributions to both triangle and box form factors in a
private version of the public code {\tt HPAIR}\cite{HPAIR}, which
computes the NLO-QCD cross section for Higgs pair production in the SM
and in the MSSM. We find that the two-loop squark contributions can
have a non-negligible effect in scenarios with stop masses below the
TeV scale. We conclude by discussing the limitations of the
approximation of vanishing external momenta. Finally, in the
appendices we collect some analytic formulae for the two-loop box form
factors, and we show how our results can be adapted to the case of
Higgs pair production in the next-to-minimal supersymmetric extension
of the SM (NMSSM).

\section{Higgs pair production via gluon fusion 
at NLO in the MSSM }
\label{sec:general}

In this section we summarize some general results on the gluon-fusion
production of a pair of neutral Higgs scalars, denoted as $\phi$ and
$\chi$ (each of them can be either $h$ or $H$). The hadronic cross
section for the process $h_1 + h_2 \to \phi+\chi + X$ at
center-of-mass energy $\sqrt{s}$ can be written as
\bea
\invm
\frac{d \sigma}{d \invm}
&= & 
\sum_{a,b}\int_0^1 dx_1 dx_2 \,\,f_{a,h_1}(x_1,\muF)\,
f_{b,h_2}(x_2,\muF)\,
\int_0^1 dz~ \delta \left(z-\frac{\invm}{\hat s} \right)
\invm
\frac{d \hat\sigma_{ab}}{d  \invm}~,
\label{sigmafull}
\eea
where: $\invm$ is the invariant mass of the $\phi+\chi$ system;
$f_{a,h_i}(x,\muF)$ is the density for the parton of type $a$ (with $a
= g,q,\bar{q}$) in the colliding hadron $h_i$; $\muF$ is the
factorization scale; $\hat{s}=s\,x_1 \,x_2$ is the partonic
center-of-mass energy; $\hat\sigma_{ab}$ is the cross section for
the partonic subprocess $ ab \to \phi + \chi+X$.  The partonic cross
section can be written in terms of the LO contribution
$\sigma^{(0)}_{\phi\chi}$ as
\be
\invm\,\frac{d \hat\sigma_{ab}}{d \invm}
~=~
\sigma^{(0)}_{\phi\chi}\,z \, G_{ab}(z)~.
\label{Geq}
\ee
The LO cross section is
\beq
\sigma^{(0)}_{\phi\chi} = \frac1{1 + \delta_{\phi\chi}} 
\frac{G_F^2\,\alpha_s^2 (\muR)}{256 \,(2\pi)^3 } 
\int_{\hat{t}_-}^{\hat{t}_+} d\hat{t} \,
\left(\left|{\cal F}^{\phi\chi,\,1\ell}\right|^2
+ \left|{\cal G}^{\phi\chi,\,1\ell}\right|^2\right)~,
 \label{sigmalo}
\eeq
where: $G_F$ is the Fermi constant; $\alpha_s(\muR)$ is the strong
gauge coupling expressed in the $\overline{\rm MS}$ renormalization
scheme at the scale $\muR$; the Mandelstam variables of the partonic
process, $\hat{t}$ and (for later convenience) $\hat{u}$, are
defined as
\beq
\ \quad \hat{t} &=& -\f{1}{2}\left(\szhat-m_{\phi}^2-m_{\chi}^2 - 
\cos\theta \sqrt{\lambda(\szhat, m_{\phi}^2, m_{\chi}^2)}\right)~,\\
\quad \hat{u} &=& -\f{1}{2} \left(\szhat-m_{\phi}^2-m_{\chi}^2 + 
\cos\theta \sqrt{\lambda(\szhat, m_{\phi}^2, m_{\chi}^2)}\right)~, 
\label{mandelstam}
\eeq
with $\theta$ the scattering angle in the partonic center-of-mass
system, and
\beq
\lambda(x,y,z)=(x-y-z)^2-4 y z~.
\eeq
The integration limits in \Eqref{sigmalo} are given by
\beq
\hat{t}_\pm = -\f{1}{2}\left(\szhat-m_{\phi}^2-m_{\chi}^2 \mp 
\sqrt{\lambda(\szhat, m_{\phi}^2, m_{\chi}^2)}\right)~,
\label{tplusminus}
\eeq 
corresponding to $\cos\theta = \pm 1$. 
Finally, in \Eqref{sigmalo} ${\cal F}^{\phi\chi,\,1\ell}$ and
${\cal G}^{\phi\chi,\,1\ell}$ represent the one-loop parts of the
spin-zero and spin-two form factors for the process $gg\rightarrow
\phi\chi$, respectively. While the spin-two form factor ${\cal
  G}^{\phi\chi}$ receives only contributions from box diagrams, the
spin-zero form factor ${\cal F}^{\phi\chi}$ can be decomposed in box
and triangle contributions as:
\beq
\label{boxtriangle}
{\cal F}^{\phi\chi} ~=~
F_{\Box}^{\phi\chi}  ~+~\cupc\,F_{\Delta}^{h} ~+~  \cdpc\,F_{\Delta}^{H}~.
\eeq
In particular, $F_{\Box}^{\phi\chi}$ contains the spin-zero part of
the box diagrams, while $F_{\Delta}^{h}\, (F_{\Delta}^{H})$ contains
the contribution of the triangle diagrams for the production of an
off-shell scalar $h\, (H)$ which subsequently decays into the pair
$\phi \chi$ through the factor $\cupc\, (\cdpc)$, defined as
\beq
\cupc~=~\lambda_{h \phi \chi}\,\f{\mz^2}{\szhat-m_{h}^2+
i \,m_{h}\,\Gamma_{h}} ~,
\label{c2tria}
\eeq
where $\lambda_{h \phi \chi}$ is the trilinear scalar
coupling\,\footnote{We normalize all trilinear Higgs couplings to
  $\lambda_0 = \mz^2/v$, with $v= (\sqrt2 G_F)^{-1/2} \approx 246
  \text{ GeV}$.} and $\Gamma_{h}$ is the width of the scalar $h$ (in
turn, $\cdpc$ is obtained from \Eqref{c2tria} with the replacement $h
\to H$).
The form factor $F_{\Delta}^\phi$ is decomposed in one- and two-loop
parts as
\be
F_{\Delta}^{\phi} ~=~ F_{\Delta}^{\phi,\,1\ell}
~+~ \frac{\alpha_s}{\pi} \, F_{\Delta} ^{\phi,\,2\ell}
~+~{\cal O}(\alpha_s^2)~,
\label{Fdec}
\ee
and analogous decompositions hold for $F_{\Box}^{\phi\chi}$, ${\cal
  F}^{\phi\chi}$ and ${\cal G}^{\phi\chi}$.

\vfill
\newpage

The coefficient function $G_{ab}(z)$ in \Eqref{Geq} can in turn be
decomposed, up to NLO terms, as
\be
G_{a b}(z)  ~=~  G_{a b}^{(0)}(z) 
~+~ \frac{\alpha_s}{\pi} \, G_{a b}^{(1)}(z) ~+~{\cal O}(\alpha_s^2)\, ,
\label{Gdec}
\ee
with the LO contribution given only by the gluon-fusion channel:
\bea
G_{a b}^{(0)}(z) & = & \delta(1-z) \,\delta_{ag}\, \delta_{bg} \, .
\eea
The NLO terms include, besides the $gg$ channel, also the one-loop
induced processes $gq \rightarrow q \phi \chi$ and 
$q \bar{q} \rightarrow g \phi \chi$.
The $gg$-channel contribution, involving two-loop
virtual corrections to $g g \rightarrow \phi \chi$ and one-loop real
corrections from $ gg \to \phi \chi g$, can be written as 
\bea
G_{g g}^{(1)}(z) & = & \delta(1-z) \left[C_A \, \frac{~\pi^2}3 
 \,+ \beta_0 \, \ln \left( \frac{\muR^2}{\muF^2} \right) \,+ 
\,\f{\int_{t_-}^{t_+} d\hat{t}\,\left( 
{\cal C}_{\scriptscriptstyle{\rm NLO}}^{\phi\chi} 
\, +  \,{\rm h.c.}\right)  }{\int_{\hat{t}_-}^{\hat{t}_+} d\hat{t}\, 
\left(\left|{\cal F}^{\phi\chi,\,1\ell}\right|^2
+ \left|{\cal G}^{\phi\chi,\,1\ell}\right|^2\right)}
  \right]  \nn \\[1mm]
&+ & 
  P_{gg} (z)\,\ln \left( \frac{\hat{s}}{\muF^2}\right) +
    C_A\, \frac4z \,(1-z+z^2)^2 \,{\cal D}_1(z) +  C_A\, {\cal R}_{gg}  \, , 
\label{real}
\eea
where 
\be
\label{nloC}
{\cal C}_{\scriptscriptstyle{\rm NLO}}^{\phi\chi} ~=~
\left({\cal F}^{\phi\chi,\,1\ell}\right)^*\,
\left({\cal F}^{\phi\chi,\,2\ell} ~+~
{\cal F}_{\scriptscriptstyle \Delta\Delta}^{\phi \chi}\,\right)
~+~ 
\left({\cal G}^{\phi\chi,\,1\ell}\right)^*\,
\left({\cal G}^{\phi\chi,\,2\ell} ~+~
{\cal G}_{\scriptscriptstyle \Delta\Delta}^{\phi \chi}\,\right) ~.
\ee

In \Eqref{real}, $C_A =N_c$ ($N_c$ being the number of colors),
$\beta_0 = (11\, C_A - 2\, N_f)/6 $ ($N_f$ being the number of active
flavors) is the one-loop $\beta$-function of the strong coupling in
the SM, $P_{gg}$ is the LO Altarelli-Parisi splitting function
\be
P_{gg} (z) ~=~2\,  C_A\,\left[ {\cal D}_0(z) +\frac1z -2 + z(1-z) \right]
\label{Pgg} \, ,
\ee 
and
\be
{\cal D}_i (z) =  \left[ \frac{\ln^i (1-z)}{1-z} \right]_+  \label {Dfun} \, .
\ee

The first line of \Eqref{real} displays the two-loop virtual
contribution regularized by the infrared singular part of the
real-emission cross section.  The second line contains the
non-singular contribution from the real gluon emission in the
gluon-fusion process.  The function ${\cal R}_{gg}$ is obtained from
one-loop diagrams where only quarks or squarks circulate into the
loop, and in the limit of vanishing external momenta it becomes ${\cal
  R}_{gg} \to -11 (1-z)^3/(6 z)$. The form factors ${\cal
  F}_{\scriptscriptstyle \Delta\Delta}^{\phi \chi}$ and ${\cal
  G}_{\scriptscriptstyle \Delta\Delta}^{\phi \chi}$ in \Eqref{nloC}
represent the contributions of two-loop double-triangle diagrams with
$t/u$-channel gluon exchange. In the limit of vanishing external
momenta, the double-triangle form factors can be expressed in terms of
the one-loop triangle form factors:
\be
\label{doubletriangle}
{\cal F}_{\scriptscriptstyle \Delta\Delta}^{\phi \chi}
~~\underset{p_i=0}{\longrightarrow}~~   
\frac12\,F_{\Delta}^{\phi,1\ell}\,F_{\Delta}^{\chi,1\ell}~,
~~~~~~~~~~~
{\cal G}_{\scriptscriptstyle \Delta\Delta}^{\phi \chi}
~~\underset{p_i=0}{\longrightarrow}~~  
-\f{p_{\scriptscriptstyle T}^2}{4 \,\hat{t}\hat{u}}\,
(\szhat-m_{\phi}^2-m_{\chi}^2) \, 
F_{\Delta}^{\phi,1\ell}\,F_{\Delta}^{\chi,1\ell}~,
\ee
 with
\beq
p_{\scriptscriptstyle T}^2~=~
\frac{\left( \hat{t}-m_{\phi}^2\right)\left( \hat{u}-m_{\phi}^2\right)}
{\szhat}~-~m_{\phi}^2~.
\eeq

Finally, the contributions of the $gq \rightarrow q \phi \chi$ and $q
\bar{q} \rightarrow g \phi \chi$ channels are given by:
\be
G_{q \bar{q}}^{(1)}(z) ~=~   {\cal R}_{q \bar{q}} \, , ~~~~~~~~~~~
G_{q g}^{(1)}(z) ~=~  P_{gq}(z) \left[ \ln(1-z) + 
 \frac12 \ln \left( \frac{\hat{s}}{\muF^2}\right) \right] + {\cal R}_{qg} \,,
\label{qqqg}
\ee
where
\be
P_{gq} (z) ~=~  C_F \,\frac{1 + (1-z)^2}z~, 
\ee
with $C_F = (N_c^2-1)/(2\,N_c)$. The functions ${\cal R}_{q \bar q}$
and ${\cal R}_{q g}$ in \eqref{qqqg} are obtained from one-loop quark
and squark diagrams, and in the limit of vanishing external momenta
become ${\cal R}_{q \bar q} \to 32 \,(1-z)^3/(27 z)$,
${\cal R}_{q g} \to 2\,z/3 - (1-z)^2/z$.


\section{Box form factors in the limit of vanishing external 
momenta}
\label{sec:formfactors}

As mentioned in section~\ref{sec:intro}, exact results for the
one-loop form factors ${\cal F}^{\phi\chi,\,1\ell}$ and ${\cal
  G}^{\phi\chi,\,1\ell}$ which determine the cross section for Higgs
pair production at the LO have been known for a long time, both for
the SM~\cite{Glover:1987nx} and for the MSSM~\cite{Plehn:1996wb,
  Belyaev:1999mx, BarrientosBendezu:2001di}. At two loops, the
triangle contributions to the form factors can be borrowed from the
calculation of the cross section for single Higgs
production. However, explicit formulae for the contributions of
triangle diagrams involving quarks, squarks and gluinos are available
only in approximate form, assuming the existence of some hierarchy
among the relevant masses and momenta~\cite{Harlander:2003bb,
  Harlander:2004tp, Degrassi:2008zj, Degrassi:2010eu,
  Harlander:2010wr, Degrassi:2012vt}. Two-loop results for the box
contributions to the form factors are known only for the diagrams
involving top quarks and gluons, and    only in the heavy-top
limit~\cite{Dawson:1998py}.

In this section we present a novel calculation of the contributions of
diagrams involving top quarks and stop squarks to the box component
$F_{\Box}^{\phi\chi}$ of the spin-zero form factor ${\cal
  F}^{\phi\chi}$, up to the two-loop order. We restrict our
calculation to the limit of vanishing external momenta, which, for the
top-gluon contribution alone, corresponds to the heavy-top limit. Note
that the corresponding triangle component $F_{\Delta}^{\phi}$ can be
extracted from ref.~\cite{Degrassi:2008zj}, and that the spin-two form
factor ${\cal G}^{\phi\chi}$ vanishes in the zero-momentum limit.  We
also present results for the contributions of the diagrams involving
sbottom squarks, under the additional approximation of vanishing
bottom mass. Finally, we show how the formulae for the two-loop part
of the form factors are affected by a change in the renormalization
scheme of the parameters entering the one-loop part.

It is convenient to decompose the triangle and box form factors for
the production of scalar mass eigenstates as
\bea
\label{Fdeltah}
F_{\Delta}^{h}&=& -T_F\,\left[
-\sin\alpha \, {\cal H}_{1} +\cos\alpha \, {\cal H}_{2}\right]~,\\[1mm]
F_{\Delta}^{\smallH}&=& -T_F\,\left[
~~\cos\alpha \, {\cal H}_{1} +\sin\alpha \, {\cal H}_{2}\right]~,\\[1mm]
F_{\Box}^{hh}&=& -T_F\,\left[
\sin^2\alpha \, {\cal H}_{11}
+\cos^2\alpha \, {\cal H}_{22}
-2\,\sin\alpha\, \cos\alpha\,{\cal H}_{12}\right]~,\\[1mm]
F_{\Box}^{\smallH\smallH}\!&=& -T_F\,\left[
\cos^2\alpha \, {\cal H}_{11}
+\sin^2\alpha \, {\cal H}_{22}
+2\,\sin\alpha\, \cos\alpha\, {\cal H}_{12}\right]~,\\[1mm]
F_{\Box}^{h\smallH}&=& -T_F\,\left[
(\cos^2\alpha-\sin^2\alpha) \, {\cal H}_{12}
-\sin\alpha\, \cos\alpha\,({\cal H}_{11}-{\cal H}_{22})\right]~,
\label{FBoxhH}
\eea
where $T_F=1/2$ is a color factor (we make it explicit to follow the
notation of ref.~\cite{Degrassi:2008zj}), the angle $\alpha$ relates
the scalar mass eigenstates, $h$ and $H$, to the real parts of the
neutral components of the two MSSM Higgs doublets, $S_1$ and $S_2$,
\beq
\left(\!\!\begin{array}{c} H\\h\end{array}\!\!\right) ~=~
\left(\!\begin{array}{cc} \cos\alpha&\sin\alpha\\
-\sin\alpha&\cos\alpha\end{array}\!\right)\,
\left(\!\!\begin{array}{c} S_1\\S_2\end{array}\!\!\right)~,
\eeq
and ${\cal H}_{i}$ and ${\cal H}_{ij}$, with $i,j = (1,2)$, are form
factors in the interaction basis. As mentioned above, the form factors
${\cal H}_{i}$ were computed in refs.~\cite{Degrassi:2008zj,
  Degrassi:2010eu, Degrassi:2012vt} for single Higgs
production. Finally, we further decompose the form factors ${\cal
  H}_{ij}$ into top/stop and bottom/sbottom contributions, ${\cal
  H}_{ij} = {\cal H}^t_{ij}\,+\,{\cal H}^b_{ij}$.

\subsection{Top/stop contributions via the low-energy theorem}
\label{sec:LET}

In our derivation of the top/stop contributions to the box form
factors we rely on the same LET for Higgs
interactions~\cite{Ellis:1975ap, Shifman:1979eb, Kniehl:1995tn} that
was employed in ref.~\cite{Degrassi:2008zj} for the calculation of the
top/stop contribution to the triangle form factors. In our case, the
LET connects the form factor for the interactions of two gluons with
two Higgs scalars at vanishing external momenta to the second
derivatives of the gluon self-energy with respect to the Higgs
scalars. In particular, we can write the top/stop contributions to the
form factors in the interaction basis as
\beq
{\cal H}^t_{ij} ~=~ \frac{2\pi\,v^2}{\alpha_s\,T_F}~
\frac{~\partial \Pi^t(0)}{\partial S_i\,\partial S_j}~,
\eeq
where $\Pi^t(q^2)$ denotes the top/stop contribution to the transverse
part of the dimensionless (i.e., divided by $q^2$) self-energy of the
gluon. In analogy with the effective-potential calculation of the MSSM
Higgs masses in ref.~\cite{Degrassi:2001yf} and with the LET
calculation of single Higgs production in ref.~\cite{Degrassi:2008zj},
the dependence of the gluon self-energy on the Higgs fields $S_i$ can
be identified through the field dependence of the top mass $m_t$, the
stop masses $\tu$ and $\td$ and the stop mixing angle $\theta_t$,
defined as
\beq
\left(\!\!\begin{array}{c} \tilde t_1\\\tilde t_2\end{array}\!\!\right) ~=~
\left(\!\begin{array}{cc} \cos\theta_t&\sin\theta_t\\
-\sin\theta_t&\cos\theta_t\end{array}\!\right)\,
\left(\!\!\begin{array}{c} \tilde t_L\\\tilde t_R\end{array}\!\!\right)~.
\eeq
A lengthy but straightforward application of the chain rule for the
derivatives allows us to express the form factors as
\bea
{\cal H}^t_{11}& = & \frac{2\,\mt^2}{\sin^2\beta}\,\left[
\frac{1}{2} \, \mu^2 \,\sdt^2  \,F_3 
~+~ \frac{ \mu^2}{\tu-\td}\, F\right]\nn\\[1mm]
&+&4\,\mz^2\,\left[ \mt\,\mu\,\cot\beta\,\sdt\,\widetilde F_2
~+~ \mz^2\,\cos^2\beta\,\widetilde F_3
~+~ \frac12\,D\right]\,,
\label{H11} \\[3mm]
{\cal H}^t_{12} & = & \frac{2\,\mt^2}{\sin^2\beta}\,\left[
\mu\, m_t\, \sdt \,  F_2 
~+~ \frac{1}{2}\,  \mu \,A_t \, \sdt^2 \, F_3
~+~ \frac{\mu\,A_t}{\tu-\td}\, F\right] \nn \\[1mm]
&+&4\,\mz^2\,\left[  \mt^2\,\cot\beta\,\widetilde F_1 ~+~
\frac12\,\mt\,(A_t\,\cot\beta-\mu)\,\sdt\,\widetilde F_2
~-~ \mz^2\,\sin\beta\,\cos\beta\,\widetilde F_3 \right]\,, 
\label{H12} \\[3mm]
{\cal H}^t_{22} & = &\frac{2\,\mt^2}{\sin^2\beta}\,\left[
2\,  m_t^2\, F_1
~+~ 2\, m_t\,  A_t\, \sdt\, F_2 
~+~ \frac{1}{2}\,  A_t^2\, \sdt^2\, F_3 
~+~ \frac{A_t^2}{\tu-\td}\, F ~+~ G\right]\nn\\[1mm]
&+&4\,\mz^2\,\left[-~2\,\mt^2\,\widetilde F_1
-\mt\,A_t\,\sdt\,\widetilde F_2
~+~ \mz^2\,\sin^2\beta\,\widetilde F_3
~-~ \frac12\,D\right]~,
\label{H22}
\eea 
where $A_t$ is the trilinear soft-SUSY breaking Higgs-stop coupling,
$\mu$ is the Higgs/higgsino mass term in the superpotential (with the
sign convention of refs.~\cite{Degrassi:2008zj,Degrassi:2001yf}), and
we define $\sdt\equiv\sin 2\theta_t$ and, for later convenience,
$\cdt\equiv\cos 2\theta_t$.  We note that the first line of each
equation contains contributions from diagrams in which the Higgs
scalars interact only via the top Yukawa coupling, whereas the second
line contains sub-dominant contributions from diagrams in which one or
both Higgs scalars interact with the squarks via a $D$-term induced EW
coupling. The functions $F_i$, $F$, $G$, $\widetilde F_i$ and $D$ are
combinations of the first and second derivatives of the gluon
self-energy with respect to the parameters $\mt^2$, $\tu$, $\td$ and
$\cdtt^2$. At one loop, the functions in the first lines of
eqs.~(\ref{H11})--(\ref{H22}) read
\beq
\label{Fi1loop}
F_1^{1\ell} = \frac16
\left(\frac{1}{\tuq}+\frac{1}{\tdq}+\frac{4}{\mt^4}\right),~~~\,
F_2^{1\ell} = \frac16
\left(\frac{1}{\tuq}-\frac{1}{\tdq}\right),~~~\,
F_3^{1\ell} = \frac16
\left(\frac{1}{\tuq}+\frac{1}{\tdq}
-\frac{2}{\tu\,\td}\right),
\eeq
\vspace*{-3mm}
\beq
\label{FG1loop}
F^{1\ell}~=~ -\frac16
\left(\frac{1}{\tu}-\frac{1}{\td}\right),~~~~~
G^{1\ell}~=~  -\frac16
\left(\frac{1}{\tu}+\frac{1}{\td}+\frac{4}{\mt^2}\right),
\eeq
and those in the second lines read
\bea
\widetilde F_1^{1\ell} &=& 
\frac{\dtl+\dtr}{12}\,\left(\frac1\tuq +\frac1\tdq\right)
~+~\cdt\,\frac{\dtl-\dtr}{12}\,\left(\frac1\tuq -\frac1\tdq\right)
~,\\[2mm]
\label{F2t1loop}
\widetilde F_2^{1\ell} &=& 
\frac{\dtl+\dtr}{12}\,\left(\frac1\tuq -\frac1\tdq\right)
~+~\cdt\,\frac{\dtl-\dtr}{12}\,\frac{(\tu-\td)^2}{\tuq\,\tdq}~,\\[2mm]
\widetilde F_3^{1\ell} &=&
\frac{(\dtl)^2+(\dtr)^2}{12}\,\left(\frac1\tuq +\frac1\tdq\right)
~-~\sdt^2\,\frac{(\dtl-\dtr)^2}{24}\,\frac{(\tu-\td)^2}{\tuq\,\tdq}\nn\\[1mm]
&&~+~\cdt\,\frac{(\dtl)^2-(\dtr)^2}{12}\,\left(\frac1\tuq -\frac1\tdq\right)
~,\\[2mm]
\label{D1loop}
D^{1\ell} &=& 
-\frac{\dtl+\dtr}{12}\,\left(\frac1\tu +\frac1\td\right)
~-~\cdt\,\frac{\dtl-\dtr}{12}\,\left(\frac1\tu -\frac1\td\right)
~,
\eea
where 
\be
\dtl = \frac12-\frac23\,\sin^2\theta_\smallw~,~~~~~
\dtr = \frac23\,\sin^2\theta_\smallw~,
\ee
$\theta_\smallw$ being the Weinberg angle.

In appendix~\ref{app:functions} we provide the explicit definitions of
the two-loop functions $F_i^{2\ell}$, $F^{2\ell}$, $G^{2\ell}$,
$\widetilde F_i^{2\ell}$ and $D^{2\ell}$ in terms of the derivatives
of the gluon self-energy. For the latter, we define the shortcut $Z
\,\equiv\, (2/T_F)\;\Pi^{2\ell,\, t}(0)$, after decomposing the gluon
self-energy in one- and two-loop parts as
\beq
\Pi(q^2) ~=~ \frac{\as}{\pi} \,\Pi^{1\ell}(q^2)
~+~\left(\frac{\as}{\pi}\right)^2
\,\Pi^{2\ell}(q^2)~+~{\cal O}(\as^3)~.
\eeq
Analytic formulae for the first derivatives of $Z$, computed under
the assumption that the one-loop part of the gluon self-energy is
expressed in terms of $\drbar$-renormalized top/stop parameters, were
given in ref.~\cite{Degrassi:2008zj}. Indeed, the functions $F$, $G$
and $D$ entering eqs.~(\ref{H11})--(\ref{H22}) coincide with those
defined in that paper for the case of single Higgs
production. Analytic formulae for the second derivatives of $Z$, which
enter the functions $F_i$ and $\widetilde F_i\,$, can be easily
obtained from those for the first derivatives, using the recursive
relations for the derivatives of the two-loop function
$\Phi(m_1^2,m_2^2,m_3^2)$ given e.g.~in appendix~A of
ref.~\cite{Dedes:2002dy}. However, those formulae are too lengthy to
be given explicitly in print, thus we make our results available upon
request as a fortran routine.

\subsection{Bottom/sbottom contributions for vanishing bottom mass}
\label{sec:sbottom}
The LET employed in the previous section to compute the top/stop
contributions to the box form factors relies on the assumption that
the external momenta are negligible with respect to the masses of all
particles running in the loops. Obviously, this assumption cannot hold
for the contributions involving bottom quarks, nor for those involving
quarks of the first two generations. In ref.~\cite{Degrassi:2010eu}
the bottom/sbottom contributions to single Higgs production were
computed with an asymptotic expansion in the heavy supersymmetric
masses (which we collectively denote by $\msusy$), up to terms that
induce ${\cal O}(\mb^2/m_\phi^2)$, ${\cal O}(\mb/\msusy)$ and
${\cal O}(\mz^2/\msusy^2)$ contributions to the triangle form
factors. In the calculation of the bottom/sbottom contributions to the
box form factors we follow the same approach as in
ref.~\cite{Degrassi:2010eu}, but we make for simplicity the further
approximation that the bottom mass and the left-right mixing in the
sbottom mass matrix are set to zero (i.e., $\mb=\theta_b=0$),
effectively killing the Yukawa-induced interactions between Higgs
bosons and bottom (s)quarks.\footnote{Since the sbottom mixing
  contains a $\tb$-enhanced term, this might not be a good
  approximation at large $\tb$.} This leaves us with the contributions
of diagrams in which the Higgs bosons interact with the squarks $\sbl$
and $\sbr$ only via $D$-term induced EW couplings, which are
parametrically of the same order as the terms involving the functions
$\widetilde F_3$ and $D$ in the top/stop contributions,
eqs.~(\ref{H11})--(\ref{H22}). In particular, we find
\bea
\left.{\cal H}^b_{11}\right|_{D{\rm \mbox{-}term}}
& = & 4\,\mz^4\,\cos^2\beta\,\widetilde F_{3\,b}
~+~ 2\,\mz^2\,D_b~,
\label{H11b} \\[3mm]
\left.{\cal H}^b_{12}\right|_{D{\rm \mbox{-}term}}
& = &-~  4\,\mz^4\sin\beta\,\cos\beta\,\widetilde F_{3\,b}~, 
\label{H12b} \\[3mm]
\left.{\cal H}^b_{22}\right|_{D{\rm \mbox{-}term}} 
& = &  4\,\mz^4\,\sin^2\beta\,\widetilde F_{3\,b}
~-~ 2\,\mz^2\,D_b~.
\label{H22b}
\eea 
The one-loop parts of the functions $\widetilde F_{3\,b}$ and $D_b$
read, in this approximation, 
\beq
\widetilde F_{3\,b}^{1\ell}~=~
\frac{(\dbl)^2}{6\,\blq}~+~\frac{(\dbr)^2}{6\,\brq}~,~~~~~~~~~
D_b^{1\ell}~=~
-\frac{\dbl}{6\,\bl}~-~\frac{\dbr}{6\,\br}~,
\eeq
where 
\be
\dbl = -\frac12+\frac13\,\sin^2\theta_\smallw~,~~~~~
\dbr = -\frac13\,\sin^2\theta_\smallw~.
\ee

We obtained the two-loop parts of the functions $\widetilde F_{3\,b}$
and $D_b$ by explicit computation of the relevant two-loop diagrams,
setting $\mb=\theta_b=0$ from the start and taking the first
non-vanishing term of an asymptotic expansion in the heavy
superparticle masses (for an outline of this approach, see section 3
of ref.~\cite{Degrassi:2010eu}). Under the assumption that the
one-loop parts of the form factors are expressed in terms of
$\drbar$-renormalized sbottom masses at the scale $Q$, we get
\bea 
\widetilde F_{3\,b}^{2\ell} &=&
(\dbl)^2\,\Bigg[\frac{C_F}{12 \,m_{\tilde{g}}^4}
\left(\frac{-4+17\, \xl-29 \,\xl^2+19\,\xl^3
    -3 \,\xl^4}{(1-\xl)^3\,\xl^3}
  +\frac{4}{\xl^3}\ln \frac{m_{\tilde{g}}^2}{Q^2}
  -\frac{4}{(1-\xl)^3}\ln \xl\right) \nonumber \\
&+& \frac{C_A}{12 \,m_{\tilde{g}}^4} \left(\frac{1-3
    \,\xl}{(1-\xl)^2\, \xl^2}
  -\frac{2}{(1-\xl)^3}\ln \xl\right)\Bigg]
+~ (\smalll \rightarrow \smallr)~,\\[3mm]
D_{b}^{2\ell} &=& \dbl\, \Bigg[-\frac{C_F}{12 \,m_{\tilde{g}}^2}
\left( \frac{-2+9 \,\xl-10 \,\xl^2
+3 \,\xl^3}{(1-\xl)^2\,\xl^2}
+\frac{2}{\xl^2} \ln \frac{m_{\tilde{g}}^2}{Q^2}
+ \frac{2}{(1-\xl)^2}\ln \xl\right) \nonumber \\
&-&\frac{C_A}{12 \,m_{\tilde{g}}^2}\left(
  \frac{1}{(1-\xl)\,\xl}+\frac{1}{(1-\xl)^2}\ln
  \xl\right)\Bigg] +~ (\smalll \rightarrow \smallr)~, 
\eea
with
$x_{\smalll,\smallr}=m_{\tilde{b}_{\smalll,\smallr}}^2/m_{\tilde{g}}^2$
and the notation $(\smalll \rightarrow \smallr)$ means a term that is
obtained from the previous one with the exchanges $\xl \to x_\smallr$
and $\dbl \to \dbr$.  We find that, when $\mb=\theta_b=0$, there are
no infrared-divergent parts in the two-loop bottom/sbottom diagrams,
therefore our results could also be obtained as the first
non-vanishing term of a Taylor expansion of those diagrams in the
external momenta. On the other hand, we stress that our results cannot
be obtained by setting $\mt=\theta_t=0$ in the LET results for the
top/stop contributions, because the latter rely on the assumption that
the external momenta are much smaller than the quark mass. Finally,
the contributions of the first two generations of quarks and squarks
can be obtained, by means of trivial substitutions, from
eqs.~(\ref{H11})--(\ref{H22}) and from the results presented in this
section.

\subsection{Change of renormalization scheme}
\label{OSshift}

The results presented in sections~\ref{sec:LET} and \ref{sec:sbottom}
were obtained under the assumption that the parameters entering the
one-loop part of the form factors are expressed in the $\drbar$
renormalization scheme. If a different scheme is used, the
two-loop part of the form factor receives a shift
\beq
{\cal H}_{ij}^{2\ell} ~\longrightarrow~
{\cal H}_{ij}^{2\ell} ~+~ \frac{\pi}{\as}\,
\delta{\cal H}_{ij}~,
\eeq
where $\delta{\cal H}_{ij}$ is a function of the shifts of all the
parameters in the one-loop part of the form factor that are subject to
$\mathcal{O}(\as)$ corrections.\footnote{For a generic parameter $x$,
  we define the shift from the $\drbar$ scheme to a generic scheme $R$
  as $x^{\drbar}=x^R+\delta x$.}

In the top/stop sector, the parameters that need shifting are the top
mass, the stop masses, the stop mixing angle and the trilinear
coupling $A_t$. In particular, the shifts of those parameters to the
on-shell (OS) scheme adopted in our numerical discussion
can be found in appendix~B of ref.~\cite{Degrassi:2001yf}.  The shifts
$\delta {\cal H}^t_{ij}$ can then be written as
\bea
\label{dH11}
\delta {\cal H}^t_{11}& = &\frac{2\,\mt^2}{\sin^2\beta}\,\left[
\frac{1}{2} \, \mu^2 \,\sdt^2  \,\delta F_3 
~+~ \frac{ \mu^2}{\tu-\td}\, \delta F\right]\nn\\[2mm]
&+& 4\,\mz^2\,\left[
\mt\,\mu\,\cot\beta\,\sdt\,\delta \widetilde F_2
~+~ \mz^2\,\cos^2\beta\,\delta \widetilde F_3
~+~ \frac12\,\delta D\right]
 \,,\\[3mm]
\label{dH12}
\delta {\cal H}^t_{12}& = &\frac{2\,\mt^2}{\sin^2\beta}\,\left[
\mu\, m_t\, \sdt \,  \delta F_2 
~+~ \frac{1}{2}\,  \mu \,A_t \, \sdt^2 \, \delta F_3
~+~ \frac{\mu\,A_t}{\tu-\td}\, \delta F \right. \nn\\
&& ~~~~~~\left.+~\frac{1}{2}\,  \mu \,\delta A_t \, \sdt^2 \,  F_3^{1 \ell}
~+~ \frac{\mu\,\delta A_t}{\tu-\td}\, F^{1 \ell}\right]\nn\\[2mm]
&+&4\,\mz^2\,\left[
\mt^2\,\cot\beta\,\delta \widetilde  F_1
~+~\frac12\,\mt\,(A_t\,\cot\beta-\mu)\,\sdt\,\delta \widetilde F_2
~-~ \mz^2\,\sin\beta\,\cos\beta\,\delta \widetilde F_3
\right.\nn\\
&& ~~~~~~\left.
+ \frac12\,\mt\,\delta A_t\,\cot\beta\,\sdt\,\widetilde F_2^{1\ell}
\right]\,,\\[3mm]
\label{dH22}
\delta {\cal H}^t_{22} & = &\frac{2\,\mt^2}{\sin^2\beta}\,\left[
2\,  m_t^2\, \delta F_1
~+~ 2\,   m_t\, A_t\, \sdt\, \delta F_2 
~+~ \frac{1}{2}\,  A_t^2\, \sdt^2\, \delta F_3 
~+~ \frac{A_t^2}{\tu-\td}\, \delta F ~+~ \delta G\right. \nn\\
&& ~~~~~~\left.
+~2\,\mt\,\delta A_t\,\sdt\,F_2^{1 \ell} 
~+~ A_t\,\delta A_t\, \sdt^2\,  F_3^{1 \ell} 
~+~ \frac{2\,A_t\,\delta A_t}{\tu-\td}\,  F^{1 \ell}\right]\nn\\[2mm]
&+& 4\,\mz^2\,\left[
-\,2\,\mt^2\,\delta \widetilde F_1
~-~\mt\,A_t\,\sdt\,\delta \widetilde F_2
~+~ \mz^2\,\sin^2\beta\,\delta \widetilde F_3
~-~ \frac12\,\delta D\right.\nn\\
&& ~~~~~~\left.-\mt\,\delta A_t\,\sdt\,\widetilde F_2^{1\ell}\right]~,
\eea
where the one-loop parts of the functions $F_2$, $F_3$, $F$ and
$\widetilde F_2$ are given in eqs.~(\ref{Fi1loop}), (\ref{FG1loop})
and (\ref{F2t1loop}), and explicit expressions for the shifts $\delta
F_i$, $\delta F$, $\delta G$, $\delta \widetilde F_i$ and $\delta D$
are given in appendix~\ref{app:shifts}.

For what concerns the bottom/sbottom contributions, under the
approximation $\mb=\theta_b=0$ employed in section~\ref{sec:sbottom}
the shifts to the form factors reduce to
\bea
\left.\delta{\cal H}^b_{11}\right|_{D{\rm \mbox{-}term}}
& = & 4\,\mz^4\,\cos^2\beta\,\delta\widetilde F_{3\,b}
~+~ 2\,\mz^2\,\delta D_b~,
\label{dH11b} \\[3mm]
\left.\delta{\cal H}^b_{12}\right|_{D{\rm \mbox{-}term}}
& = & -~ 4\,\mz^4\sin\beta\,\cos\beta\,\delta\widetilde F_{3\,b}~, 
\label{dH12b} \\[3mm]
\left.\delta{\cal H}^b_{22}\right|_{D{\rm \mbox{-}term}} 
& = & 4\,\mz^4\,\sin^2\beta\,\delta\widetilde F_{3\,b}
~-~ 2\,\mz^2\,\delta D_b~,
\label{dH22b}
\eea 
where
\beq
\delta \widetilde F_{3\,b}~=~
-\frac{(\dbl)^2}{3\,\bls}\,\delta\bl
~-~\frac{(\dbr)^2}{3\,\brs}\,\delta\br~,~~~~~~~
\delta D_b ~=~
\frac{\dbl}{6\,\blq}\,\delta\bl
~+~\frac{\dbr}{6\,\brq}\,\delta\br~.
\eeq
If the sbottom masses in the one-loop part of the form factors are
expressed in the OS scheme, the shift $\delta\bl$ reads, for 
$\mb=\theta_b=0$,
\beq
\label{deltasbot}
\frac{\delta \bl}{\bl} ~=~
\frac{\alpha_s\,C_F}{2\pi}\left[
\ln\xl \,-\, 1 \,+\, \frac{1}{\xl}\left(2\,\ln\frac{\g}{Q^2}-3\right)
\,-\,\left(1-\frac{1}{\xl}\right)^2\ln\left|1-\xl\right|\right],
\eeq
and the shift $\delta\br/\br$ can be obtained from eq.~(\ref{deltasbot})
with the replacement $\xl\rightarrow x_\smallr$.



\section{The effect of SUSY contributions to Higgs pair production}

In this section we present numerical results for the newly-computed
SUSY contributions to the box form factors, and for their effect on
the Higgs-production cross section. We focus on the process that is
most interesting from the point of view of LHC phenomenology, i.e.~the
production of a pair of light MSSM scalars $hh$ with mass $m_h \approx
125$~GeV.

\vspace*{-1mm}
\subsection{Implementation in {\tt HPAIR}}
\label{sec:implementation}

For the numerical evaluation of the cross section, we added the
contributions of loops involving superparticles to the code {\tt
  HPAIR}~\cite{HPAIR}, whose public version includes by default the
one-loop top- and bottom-quark contributions with full mass
dependence~\cite{Plehn:1996wb} and the QCD corrections to the
top-quark contributions in the heavy-top limit~\cite{Dawson:1998py}.

For the LO cross section, we added the one-loop squark contributions
to the spin-zero and spin-two form factors, borrowing from
ref.~\cite{Belyaev:1999mx} the results with full mass dependence. At
NLO, we included our results for the two-loop stop and (partial)
sbottom contributions in the approximation of vanishing external
momenta, derived in section~\ref{sec:formfactors}. In order to improve
on that approximation, the LO cross section factored out of the
coefficient function $G_{ab}(z)$ in \Eqref{Geq} is computed with full
dependence on the top and bottom quark and squark masses. In analogy
with the implementation of the top quark loops in {\tt HPAIR}, the
$gg$-channel contribution to the NLO coefficient function in
eqs.~\eqref{real} and \eqref{nloC} -- specialized to the production of
a $hh$ pair -- becomes
 \bea
G_{g g}^{(1)}(z) & = & \delta(1-z) \left\{\,C_A \, \frac{~\pi^2}3 
 ~+~ \beta_0 \, \ln \left( \frac{\muR^2}{\muF^2} \right) ~+~ 
2\,{\rm Re}\left(\frac{{\cal F}^{hh,\,2\ell}_{\LET}}
{{\cal F}^{hh,\,1\ell}_{\LET}}\right) \right.\nonumber \\
 &+& \left.
\,\f{\int_{t_-}^{t_+} d\hat{t}~ 
{\rm Re}\left[ \left(
\left({\cal F}^{hh,\,1\ell}\right)^* 
~-~ \f{p_{\scriptscriptstyle T}^2}{2\hat{t}\hat{u}}
(\szhathh-2\,m_h^2)\,\left({\cal G}^{hh,\,1\ell}\right)^*
\right)\, 
\left({F}_{\Delta\, \LET}^{h,\, 1\ell}\right)^2\,\right]}
{\int_{\hat{t}_-}^{\hat{t}_+} d\hat{t}\, 
\left(\left|{\cal F}^{hh,\,1\ell}\right|^2
+ \left|{\cal G}^{hh,\,1\ell}\right|^2\right)}
  \right\}  \nn \\
&+ & 
  P_{gg} (z)\,\ln \left( \frac{\hat{s}}{\muF^2}\right) +
    C_A\, \frac4z \,(1-z+z^2)^2 \,{\cal D}_1(z) +  C_A\, {\cal R}_{gg}  \, , 
    \label{ggghpair}
\eea
where the subscript ``LET'' denotes form factors computed in the limit
of vanishing external momenta after setting $\mb = \theta_b = 0$. The
two-loop SUSY contributions enter the last term in the first line of
eq.~\eqref{ggghpair}, which, if only the top-quark contributions were
considered as in ref.~\cite{Dawson:1998py}, would reduce to a simple
coefficient $c_1 = 11/2$. The second line of eq.~\eqref{ggghpair}
contains the contributions of diagrams with $t/u$-channel gluon
exchange. Following ref.~\cite{Dawson:1998py}, in those contributions
we retain the full momentum dependence in the one-loop form factors
that stem from the LO matrix element, but take the limit of vanishing
external momenta, see \Eqref{doubletriangle}, in the double-triangle
form factors.  We also remark that in the NLO coefficient of
\Eqref{ggghpair} all form factors -- including those with full
momentum dependence -- are obtained for $\mb = \theta_b = 0$.

For a precise prediction of the cross section for the production of a
pair of MSSM Higgs bosons, it is essential to include the corrections
to the trilinear Higgs couplings, which can be as significant as the
corresponding corrections to the MSSM Higgs masses and mixing. Indeed,
to properly reproduce the decoupling limit in which the lightest
scalar $h$ has a SM-like self-coupling,
$\lambda^{\smallsm}_{hhh}= 3\,m_h^2/\mz^2$, the corrections to the
coupling should be computed at the same level of accuracy as the
corrections to the mass $m_h$.  The trilinear couplings are known at
one loop~\cite{Brignole:1992zv, Heinemeyer:1996tg, Barger:1991ed,
  Dobado:2002jz, Nhung:2013lpa}, but at two loops only the
$\mathcal{O}(\alpha_s\alpha_t)$ corrections have been computed, in the
effective-potential approximation, for both the
MSSM~\cite{Brucherseifer:2013qva} and the
NMSSM~\cite{Muhlleitner:2015dua}. In contrast, in this analysis we
compute the MSSM Higgs masses and mixing using the code {\tt
  FeynHiggs}~\cite{Heinemeyer:1998yj, Heinemeyer:1998np,
  Degrassi:2002fi, Frank:2006yh, Hahn:2013ria}, which includes
two-loop corrections beyond the $\mathcal{O}(\alpha_s\alpha_t)$
effective-potential ones.
Since we are anyway focusing on the effects of the SUSY contributions
to the gluon-fusion loop, we bypass the calculation of the corrections
to the trilinear couplings by relying on a simplifying approach, known
as ``hMSSM'', which was recently proposed in
refs.~\cite{Djouadi:2013vqa,Maiani:2013hud,Djouadi:2013uqa, Djouadi:2015jea}. In this
approximation one assumes that the corrections to the elements other
than $(2,2)$ of the Higgs mass matrix are negligible,
i.e.~$\Delta\mathcal{M}_{1j}^2\approx 0$ with $j=1,2$. In that case
the remaining correction $\Delta\mathcal{M}_{22}^2$, which includes
potentially large logarithmic effects from top/stop loops, can be
expressed in terms of the parameters that determine the tree-level
mass matrix (i.e.~$\tan\beta$, $\mz$ and the pseudoscalar mass $\ma$)
plus the lightest eigenvalue $m_h$, treated as an input parameter:
\beq \Delta
\mathcal{M}_{22}^2=\frac{m_h^2\,(\ma^2+\mz^2-m_h^2)-\ma^2 \mz^2 \cos^2
  2\beta}{\mz^2\cos^2\beta+ \ma^2 \sin^2\beta-m_h^2}\,.
\label{hMSSM}
\eeq 
In this approximation the trilinear couplings relevant to the
production of an $hh$ pair become
\bea
\lambda_{hhh}&=& 3\, \cos 2 \alpha \,\sin\left( \alpha+\beta \right)~+~3\,
\frac{\Delta\mathcal{M}_{22}^2}{\mz^2}\frac{\cos^3\alpha}{\sin\beta}\,,
\label{lamhMSSM}
\\ 
\lambda_{Hhh}&=&
2\, \sin 2 \alpha \,\sin\left( \alpha+\beta\right)~-~\cos 2 \alpha
\cos\left( \alpha+\beta \right)~+~3\,
\frac{\Delta\mathcal{M}_{22}^2}{\mz^2}\frac{\cos^2\alpha
  \sin\alpha}{\sin\beta}\,. 
\eea
Combining eqs.~(\ref{hMSSM}) and (\ref{lamhMSSM}) one can see that in
the decoupling limit $\ma\gg\mz$, when $\alpha \rightarrow \beta -
\pi/2$, the coupling $\lambda_{hhh}$ does indeed tend to its SM limit.
As discussed e.g.~in refs.~\cite{Bagnaschi:2015hka,Lee:2015uza}, the
approximation of neglecting the corrections
$\Delta\mathcal{M}_{1j}^2$ might not prove accurate for small $\ma$
and rather large $\mu$ and $\tan\beta$. We will therefore avoid those
choices of parameters in our numerical example.

\subsection{A numerical example}
\label{sec:numerical}

The SM parameters entering our computation of the cross section for
Higgs pair production are the $Z$ boson mass $\mz=91.1876\text{ GeV}$,
the $W$ boson mass $\mw=80.398\text{ GeV}$, the Fermi constant
$G_{F}=1.16637\cdot 10^{-5} \text{ GeV}^{-2}$ and the pole top-quark
mass $\mt=173.2\text{ GeV}$. We use the MSTW08 set of parton
distribution functions~\cite{Martin:2009iq, Martin:2009bu,
  Martin:2010db} and the associated LO and NLO values of the strong
coupling $\alpha_s$. The hadronic center-of-mass energy is set to
$\sqrt{s}= 14 \text{ TeV}$. The factorization and renormalization
scales are set to the invariant mass $\invhh$ of the Higgs boson pair.

We use the code {\tt FeynHiggs}~\cite{Heinemeyer:1998yj,
  Heinemeyer:1998np, Degrassi:2002fi, Frank:2006yh, Hahn:2013ria} to
compute the masses and mixing angle of the Higgs scalars, taking as
input the SM parameters listed above plus $\alpha_s(\mz)=0.119$. We
consider an MSSM scenario characterized by the following parameters in
the OS renormalization scheme:
\beq
&\tan \beta = 10,~~\ma=500~\text{GeV},~~\mu=-400~\text{GeV},
~~M_3=1500~\text{GeV},
\nonumber\\
&X_t=2\,\msquark~,~~
m_{\tilde{t}_\smalll}=m_{\tilde{t}_\smallr}=m_{\tilde{b}_\smallr}=\msquark, 
\label{parameters}
\eeq 
where $M_3$ denotes the soft SUSY-breaking gluino mass, we define
$X_t\equiv A_t+\mu\cot\beta$, and $m_{\tilde{t}_\smalll}$,
$m_{\tilde{t}_\smallr}$, and $m_{\tilde{b}_\smallr}$ denote the soft
SUSY-breaking masses of the third-generation squarks. We recall that,
in the OS scheme, the soft SUSY-breaking parameters in the squark
sector are defined as the parameters entering a tree-level mass matrix
that is diagonalized by the OS mixing angle (the latter defined, e.g.,
in appendix B of ref.~\cite{Degrassi:2001yf}) and has the pole squark
masses as eigenvalues. In this scheme the parameter
$m_{\tilde{b}_\smalll}$ differs from its stop counterpart
$m_{\tilde{t}_\smalll}$ by a finite
shift~\cite{Bartl:1997yd,Eberl:1999he}.

The parameters in eq.~(\ref{parameters}) -- as well as the remaining
soft SUSY-breaking parameters, which are not relevant to our
discussion -- were chosen in such a way that, for $\msquark=500$~GeV,
they reproduce the {\em light-stop} benchmark scenario proposed in
ref.~\cite{Carena:2013ytb} and studied in the context of single-Higgs
production in ref.~\cite{Bagnaschi:2014zla}. Our choices of $\ma$ and
$\tan\beta$ ensure that the lightest Higgs scalar $h$ has essentially
SM-like couplings to the top and bottom quarks, and that the
contribution of triangle diagrams with $s$-channel exchange of the
heaviest scalar $H$ is significantly suppressed, allowing us to focus
on the effects of the SUSY contributions to the box form factor.
We then vary the squark mass parameter $\msquark$ between $500$~GeV
and $1500$~GeV, which results in a lightest stop mass $\tul$ ranging
between $324$~GeV and $1326$~GeV, and in a prediction by {\tt
  FeynHiggs} for $m_h$ ranging between $122.3$~GeV and $130.7$~GeV.

\begin{figure}[t]
\centering
\vspace*{-3mm}
\input{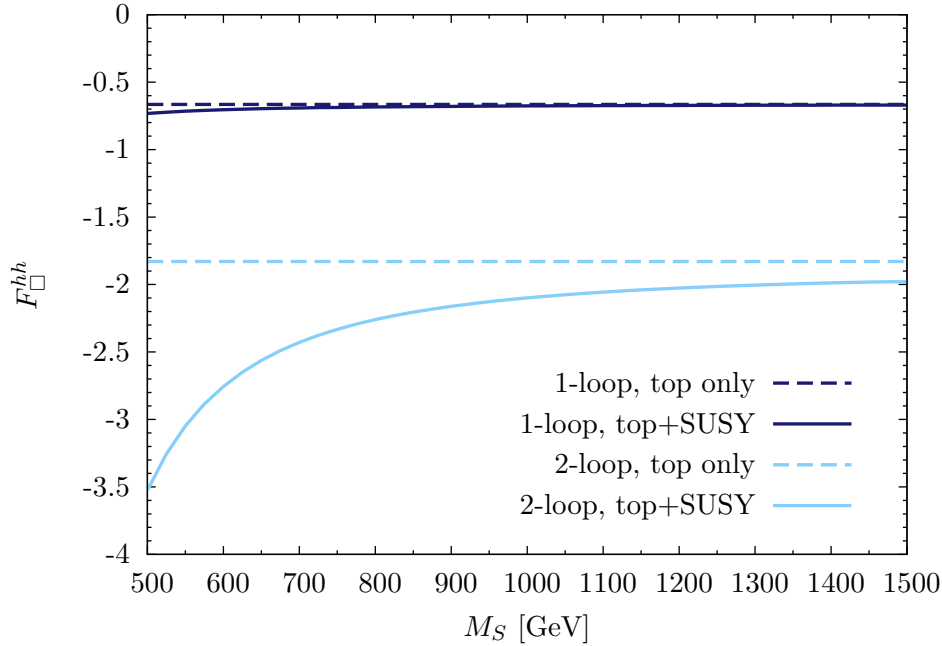}
\vspace*{-2mm}
\caption{Box form factor $F_{\Box}^{hh}$ in the vanishing-momentum limit, 
as a function of the squark-mass scale $\msquark$. Dark-blue lines 
show the one-loop form factor, light-blue lines show the two-loop form
factor. The dashed lines correspond to the top-quark contributions alone, 
whereas the solid lines include also the SUSY contributions. 
\label{fig:boxformfac}}
\end{figure}

In figure~\ref{fig:boxformfac} we plot the box form factor
$F_{\Box}^{hh}$ -- computed in the vanishing-momentum limit as
described in section~\ref{sec:formfactors} -- as a function of the
squark-mass scale $\msquark$. The solid lines correspond to the
one-loop (dark blue) and two-loop (light blue) part of the form
factor, including both the top-quark contribution and the squark
contributions.  The dashed lines correspond to the one- and two-loop
form factors including only the top contributions. The plot shows that
the squark contributions can be relevant for small squark masses, and
they are significantly larger in the two-loop form factor than in the
one-loop form factor. Moreover, the decoupling behavior of the squark
contributions for large $\msquark$ appears to be slower at two loops
than at one loop. This can be explained by the occurrence of two-loop
terms proportional to $\mt^2/\msusy^2\,\ln (\msusy^2/\mt^2)$ (with
$\msusy$ denoting generically a SUSY mass parameter), whereas at one
loop all terms decouple at least as fast as $\mt^2/\msusy^2$.

\begin{figure}[t]
\centering
\vspace*{-3mm}
\input{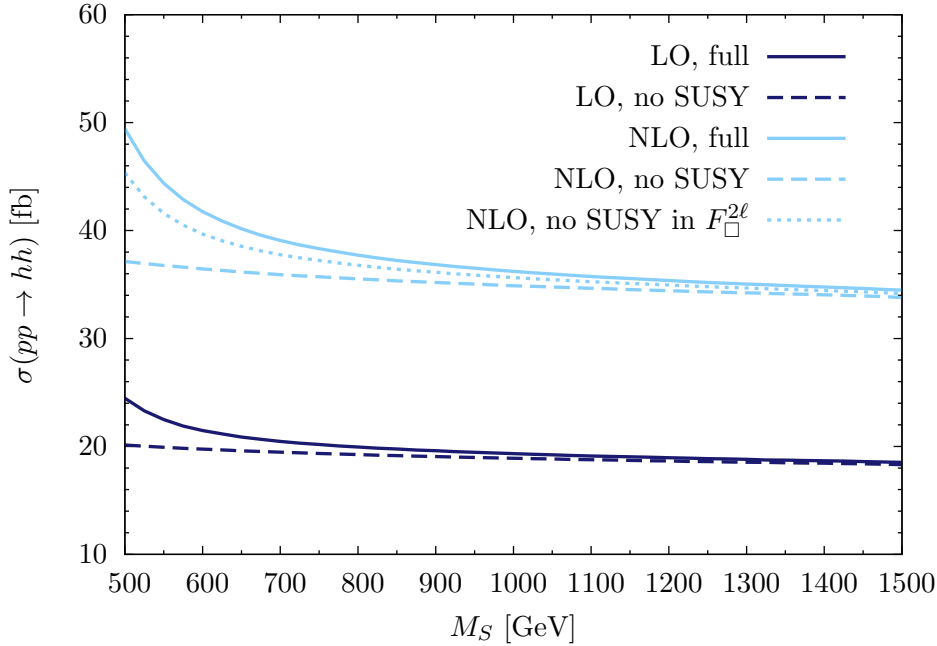}
\vspace{-2mm}
\caption{Higgs pair-production cross section
  $\sigma(pp\rightarrow hh)$ as a function of the squark-mass scale
  $\msquark$. Dark-blue lines show the LO cross section, light-blue
  lines show the NLO cross section.  The dashed lines correspond to
  the quark contributions alone, whereas the solid lines include also
  the SUSY contributions. The dotted light-blue line omits the SUSY
  contributions to the two-loop box form factor. \label{fig:can}}
\end{figure}

In figure~\ref{fig:can} we plot the cross section for the production
of a $hh$ pair as a function of $\msquark$, computed as described in
section~\ref{sec:implementation}. The dark-blue lines correspond to
the LO cross section, the light-blue lines to the NLO cross section,
and again the solid (dashed) lines correspond to form factors
including (not including) the SUSY contributions.\footnote{The mild
  $\msquark$ dependence of the dashed lines reflects the dependence of
  $m_h$ on the stop masses.} In addition, the dotted light-blue line
corresponds to the NLO cross section computed by omitting the SUSY
contributions in the two-loop part of the box form factor.
The plot shows that, for the considered choices of MSSM parameters,
the squark loops can significantly contribute to the cross section for
relatively small $\msquark$, although their effect gets quickly
suppressed when $\msquark \gsim 1$~TeV. In particular, in the {\em
  light-stop} scenario -- corresponding to the left edge of the plot
-- for our choices of $\ma$ and $\tan\beta$ the SUSY contributions
increase the NLO cross section for $h$ pair production by more than
$30\%$ (in contrast, ref.~\cite{Bagnaschi:2014zla} showed that they
reduce the cross section for the production of a single SM-like scalar
by about $20\%$). Finally, the comparison between the solid and dotted
light-blue lines shows that the newly-computed two-loop SUSY
contributions to the box form factor account for a non-negligible part
of the increase in the pair-production cross section.

\section{Discussion}

Relying on a low-energy theorem that connects the Higgs-gluon
interactions to the derivatives of the gluon self-energy, we obtained
analytic results for the contributions to Higgs pair production from
one- and two-loop box diagrams involving top quarks and stop squarks
in the limit of vanishing external momenta. We also obtained, by
direct calculation of the relevant two-loop diagrams, the subset of
bottom/sbottom contributions that involve the $D$-term-induced EW
Higgs-squark coupling and survive in the limit of vanishing bottom
mass. Combined with the existing results for the triangle diagrams in
the same approximations~\cite{Degrassi:2008zj,Degrassi:2010eu}, our
calculation allows for a consistent NLO determination of the SUSY
contributions to Higgs pair production in the MSSM.
We incorporated our results in a private version of the code {\tt
  HPAIR}, and found that the two-loop SUSY contributions to the
production of a light-scalar pair can have a non-negligible effect in
scenarios with stop masses below the TeV scale.

To conclude, a discussion is in order of the approximation of vanishing
external momenta that we employed in our calculation. Our results can
be viewed as the first term of an asymptotic expansion of the form
factor ${\cal F}^{\phi\chi,\,2\ell}$ in the heavy masses of all
particles running in the loops. Such expansion is in principle valid
only for partonic center-of-mass energies up to the lowest threshold
encountered in the relevant diagrams, which for the contributions
considered in this paper corresponds to $\sqrt{\hat{s}}= 2\,\mt$. 

In the SM, the vanishing-momentum approximation is known to work
rather well for the top-quark contributions to the production of a
single scalar $h$ with $m_h \approx 125$~GeV, because the region in
the partonic phase space with $\sqrt{\hat{s}} > 2 \, \mt$ gives only a
small contribution to the hadronic cross section. In contrast, the
same approximation is less reliable for pair production, where it is
always $\sqrt{\hat{s}} > 2\,m_h$ and the whole region up to
$\sqrt{\hat{s}} \sim 600$~GeV gives a significant contribution to the
cross section~\cite{Grigo:2013rya}. 
The factorization of the LO cross section with full momentum
dependence is expected to reduce the uncertainty of the NLO result due
to the dominance of soft and collinear gluon
effects~\cite{Dawson:1998py}.
Nevertheless, a NLO determination of the top-quark contributions to
Higgs pair production going beyond the vanishing-momentum -- or,
equivalently, infinite-top-mass -- approximation would be desirable.
Of the necessary ingredients, the contribution to
${\cal F}_\Delta^{h,\,2\ell}$ of two-loop triangle diagrams involving
top quarks and gluons is known with full top-mass dependence from
single-Higgs production; the contribution of one-loop top diagrams to
${\cal R}_{gg}, \, {\cal R}_{q \bar{q}}$ and ${\cal R}_{qg}$ is known
exactly from ref.~\cite{Maltoni:2014eza}; the contribution of
two-loop, one-particle-reducible top diagrams to
${\cal F}_{\scriptscriptstyle \Delta\Delta}^{\phi \chi}$ and
${\cal G}_{\scriptscriptstyle \Delta\Delta}^{\phi \chi}$ is relatively
easy to compute. However, an exact evaluation of the two-loop box
diagrams involving top quarks and gluons is currently not available,
and represents the bottleneck in the quest for an exact NLO
determination of the pair-production cross section.
Attempts to go beyond the limit of infinite top mass for the two-loop
box diagrams were made in refs.~\cite{Grigo:2013rya,Grigo:2015dia},
where several terms in a heavy-top asymptotic expansion of the cross
section, i.e.~terms proportional to powers of $\hat{s}/\mt^2$ or
$m_h^2/\mt^2$, were obtained. However, as shown explicitly for the LO
result in refs.~\cite{Gillioz:2012se,Dawson:2012mk}, the inclusion of
additional terms in the large-mass expansion does not necessarily
improve the evaluation of the cross section. Indeed, by including
additional terms one is improving the evaluation of the region with
$\sqrt{\hat{s}} < 2\, \mt$ at the price of worsening the evaluation of
the complementary region with $\sqrt{\hat{s}} > 2 \,\mt$, which is
approximated by a function that has the wrong behavior as ${\hat{s}}$
increases. In fact, the appropriate expansion in the region with
$\sqrt{\hat{s}} > 2\, \mt$ would be a large-momentum expansion as
opposed to a large-mass expansion.

In the MSSM, the NLO cross section for the production of a pair of
SM-like scalars $hh$ suffers from the same uncertainty as in the SM,
stemming from the incomplete knowledge of the two-loop diagrams with
top quarks and gluons. For what concerns the SUSY contributions, those
from two-loop diagrams involving squarks and gluons or quartic squark
couplings should be sufficiently well approximated, in realistic MSSM
scenarios, by the results obtained in the vanishing-momentum limit. In
contrast, some two-loop diagrams involving top, stop and gluino do
have thresholds at $\sqrt{\hat{s}}= 2\,\mt$, thus their contributions
are in principle subject to uncertainties comparable to those of the
SM contributions. The knowledge of those contributions could however
be improved following the same strategy employed in
ref.~\cite{Degrassi:2012vt} for single scalar production, namely
evaluating the top-stop-gluino box diagrams via a large-mass expansion
in the SUSY masses while treating the top quark as a light particle.

Finally, another feature specific to the MSSM calculation of $hh$
production is the possibility of large resonant contributions from
triangle diagrams with $s$-channel exchange of the heaviest scalar
$H$. In such a scenario, the determination of the NLO cross section
could be improved by using for $F_\Delta^{H,\,2\ell}$ the quark-gluon
contributions with full momentum dependence combined with the
heavy-SUSY results of refs.~\cite{Degrassi:2010eu,Degrassi:2012vt},
while retaining the vanishing-momentum approximation in
$F_\Delta^{h,\,2\ell}$ to avoid spoiling potential cancellations with
the box form factor.

\section*{Acknowledgments}
We thank Pier Paolo Giardino for useful discussions.
This work was supported in part by the Research Executive Agency (REA)
of the European Commission under the Initial Training Network
``HiggsTools'' (PITN-GA-2012-316704), and by the European Research
Council (ERC) under the Advanced Grant ``Higgs@LHC''
(ERC-2012-ADG\_20120216-321133).  The work of P.~S.~at LPTHE is
supported in part by French state funds managed by the Agence
Nationale de la Recherche (ANR), in the context of the LABEX ILP
(ANR-11-IDEX-0004-02, ANR-10-LABX-63) and of the Young Researchers
project ``HiggsAutomator'' (ANR-15-CE31-0002-01).

\vfill
\newpage

\begin{appendices}

\section{Functions entering the box form factors}
\label{app:functions}

In this appendix we provide the definitions of the functions entering
the form factors in eqs.~(\ref{H11})--(\ref{H22}) in terms of the
derivatives of the gluon self-energy. Focusing on the two-loop part of
the form factors, and defining the shortcut $Z \,\equiv\,
(2/T_F)\;\Pi^{2\ell,\, t}(0)\,$, the functions that represent the
contributions of diagrams involving only the top Yukawa coupling read
\bea
\label{defF1}
F_1^{2\ell} & = &
\DVtt + \DVtutu + \DVtdtd + 2\,\DVttu + 2\,\DVttd + 2\,\DVtutd ~,\\[3mm]
\label{defF2}
F_2^{2\ell} & = &
\DVtutu - \DVtdtd + \DVttu - \DVttd \nn\\[3mm]
&&
- \frac{4 \,\cdt^2}{\tu-\td}
\left( \DVcdtqt+\DVcdtqtu+\DVcdtqtd\right)~,\\[3mm]
\label{defF3}
F_3^{2\ell} & = &
\DVtutu + \DVtdtd -2\,\DVtutd - \frac{2}{\tu-\td}\left(\DVtu-\DVtd\right)\nn\\
&& + \frac{ 16\, \cdt^2}{(\tu-\td)^2}\,\left( \cdt^2\,\DVcdtqcdtq 
+2\,\DVcdtq \right)
- \frac{ 8\, \cdt^2}{\tu-\td}\,\left( \DVcdtqtu-\DVcdtqtd \right),\\[3mm]
\label{defF}
F^{2\ell} & = & \DVtu - \DVtd - \frac{4\,\cdt^2}{\tu-\td}\,\DVcdtq~,\\[3mm]
\label{defG}
G^{2\ell} & = &  \DVtu + \DVtd + \DVt~.
\eea

The functions that represent the sub-dominant contributions of
diagrams involving $D$-term induced EW couplings read
\bea
\widetilde F_1^{2\ell} &=&
\dtuu \,\tilde f_1  ~+~ \dtdd\,\tilde f_2
~-~4 \,{\cdt}\,{\sdt}\,\dtud\,\tilde f_3~,\\[3mm]
\widetilde F_2^{2\ell} &=&
\dtuu \,\tilde f_4  ~-~ \dtdd\,\tilde f_5
~+~2 \,\frac{\cdt}{\sdt}\,\dtud\,\tilde f_6~,\\[3mm]
\widetilde F_3^{2\ell} &=&
(\dtuu)^2 \,\DVtutu  ~+~ (\dtdd)^2\,\DVtdtd
~+~2\,\dtuu\,\dtdd\,\DVtutd\nn\\
&&~+~2\,(\dtud)^2\,\tilde f_7
~-~8\,\cdt\,\sdt\,\frac{\dtud}{\tu-\td}\,
\left(\dtuu\,\DVcdtqtu ~+~ \dtdd\,\DVcdtqtd\right)~,\\[3mm]
D_{\phantom{1}}^{2\ell} &=&
\dtuu \,\DVtu  ~+~ \dtdd\,\DVtd
~-~4 \,{\cdt}\,{\sdt}\,\frac{\dtud}{\tu-\td}\,\DVcdtq~,
\eea
where
\be
\dtuu ~=~ \frac{\dtl+\dtr}2 + \cdt\, \frac{\dtl-\dtr}2~,~~~~
\dtdd ~=~ \frac{\dtl+\dtr}2 - \cdt\, \frac{\dtl-\dtr}2~,~~~~
\dtud ~=~ -\sdt\,\frac{\dtl-\dtr}2~,
\ee
and
\bea
\tilde f_1 &=&
\DVtutu ~+~ \DVtutd ~+~ \DVttu~,\\[3mm]
\tilde f_2 &=&
\DVtdtd ~+~ \DVtutd ~+~ \DVttd~,\\[3mm]
\tilde f_3 &=&
\frac{1}{\tu-\td}\,\left(\DVcdtqtu+\DVcdtqtd+\DVcdtqt\right)~,\\[3mm]
\tilde f_4 &=&
\DVtutu ~-~ \DVtutd ~-~ \frac{4\,\cdt^2}{\tu-\td}\,\DVcdtqtu~,\\[3mm]
\tilde f_5 &=&
\DVtdtd ~-~ \DVtutd ~+~ \frac{4\,\cdt^2}{\tu-\td}\,\DVcdtqtd~,\\[3mm]
\tilde f_6 &=&
\frac{1}{\tu-\td}\,\left(\DVtu-\DVtd\right)
~-~\frac{2\,\sdt^2}{\tu-\td}\,\left(\DVcdtqtu-\DVcdtqtd\right)\nn\\[1mm]
&&~+~\frac{8}{(\tu-\td)^2}\,\left[
(1-2\,\cdt^2)\,\DVcdtq~+~\cdt^2\,\sdt^2\,\DVcdtqcdtq\right]~,\\[3mm]
\tilde f_7 &=&
\frac{1}{\tu-\td}\,\left(\DVtu-\DVtd\right)
~+~\frac{4}{(\tu-\td)^2}\,\left[
(1-4\,\cdt^2)\,\DVcdtq~+~2\,\cdt^2\,\sdt^2\,\DVcdtqcdtq\right].\nn\\[3mm]
\eea

\section{Shifts to a different renormalization scheme}
\label{app:shifts}
In this appendix we list the shifts to the functions $F_i$, $F$, $G$,
$\widetilde F_i$ and $D$ arising when the parameters $\mt$, $\ti$,
$\theta_t$ and $A_t$ in the top/stop contributions to the one-loop
part of the form factors are expressed in a renormalization scheme $R$
other than $\drbar$. Recalling the definition $x^{\drbar}=x^R+\delta
x$, the shifts to the functions read
\bea
\delta F_1 &=& 
-\frac{1}{3}\left(\frac{\delta\tu}{\tuc}~+~\frac{\delta\td}{\tdc}
~+~8\,\frac{\delta \mt}{\mt^5}\right)
~+~4\,\frac{\delta \mt}{\mt}\,F_1^{1 \ell}~,\\[2mm]
\delta F_2 &=& 
-\frac{1}{3}\left(\frac{\delta\tu}{\tuc}~-~\frac{\delta\td}{\tdc}\right)
~+~\left(3\,\frac{\delta \mt}{\mt} ~+~ \frac{\delta\sdt}{\sdt}\right)
\,F_2^{1 \ell}~,\\[2mm]
\delta F_3 &=& 
-\frac{1}{3}\left(\frac{\delta\tu}{\tuc}~+~\frac{\delta\td}{\tdc}
-\frac{\delta\tu}{\tuq\,\td}~-~\frac{\delta\td}{\tdq\,\tu}\right)
~+~\left(2\,\frac{\delta \mt}{\mt} ~+~ 2\,\frac{\delta\sdt}{\sdt}\right)
\,F_3^{1 \ell}~,\\[2mm]
\delta F &=& 
\frac{1}{6}\left(\frac{\delta\tu}{\tuq}~-~\frac{\delta\td}{\tdq}\right)
~+~\left(2\,\frac{\delta \mt}{\mt} ~-~ \frac{\delta\tu-\delta\td}{\tu-\td}
\right)\,F^{1 \ell}~,\\[2mm]
\delta G &=& 
\frac{1}{6}\left(\frac{\delta\tu}{\tuq}~+~\frac{\delta\td}{\tdq}
~+~8\,\frac{\delta \mt}{\mt^3}\right)
~+~2\,\frac{\delta \mt}{\mt}\,G^{1 \ell}~,
\eea
and
\bea
 \delta \widetilde F_1 & = & -\frac{\dtl+\dtr}{6}\,
\left(\frac{\delta\tu}\tuc +\frac{\delta\td}\tdc\right)
\,-~\frac{\dtl-\dtr}{12}\,\left[2\,\cdt\, \left(\frac{\delta\tu}\tuc
    -\frac{\delta\td}\tdc\right)
  -~\delta\cdt\,\left(\frac1\tuq -\frac1\tdq\right)~\right]\nn\\[1mm]
&& + 2\, \frac{\delta\mt}{\mt}\,\widetilde F_1^{1\ell}~,\\[5mm]
\delta \widetilde F_2 & = & -\frac{\dtl+\dtr}{6}\,
\left(\frac{\delta\tu}\tuc -\frac{\delta\td}\tdc\right)\nn\\[1mm]
&&-~\frac{\dtl-\dtr}{12}\,\left[2\,\cdt\,\left(\frac1\tu
    -\frac1\td\right) \left(\frac{\delta\tu}\tuq
    -\frac{\delta\td}\tdq\right)
  ~-~\delta\cdt\,\frac{(\tu-\td)^2}{\tuq\,\tdq}~\right]\nn\\[1mm]
&& + \left(\frac{\delta\mt}{\mt} ~+~\frac{\delta\sdt}{\sdt}\right)
\,\widetilde F_2^{1\ell}~,\\[5mm]
\delta \widetilde F_3 & = & -\frac{(\dtl)^2+(\dtr)^2}{6}\,
\left(\frac{\delta\tu}\tuc +\frac{\delta\td}\tdc\right)\nn\\[1mm]
&&-\frac{(\dtl)^2-(\dtr)^2}{12}\,\left[2\,\cdt\,
  \left(\frac{\delta\tu}\tuc -\frac{\delta\td}\tdc\right)~-~
  \delta\cdt\,\left(\frac1\tuq -\frac1\tdq\right)~\right]\nn\\[1mm]
&&+\frac{(\dtl-\dtr)^2}{12}\,\left[\sdt^2\,\left(\frac1\tu
    -\frac1\td\right) \left(\frac{\delta\tu}\tuq
    -\frac{\delta\td}\tdq\right)
  ~+~\cdt\,\delta\cdt\,\frac{(\tu-\td)^2}{\tuq\,\tdq}~\right],~\\[5mm]
\delta D & = & \frac{\dtl+\dtr}{12}\,
\left(\frac{\delta\tu}\tuq +\frac{\delta\td}\tdq\right)
\,+~\frac{\dtl-\dtr}{12}\,\left[\cdt\, \left(\frac{\delta\tu}\tuq
    -\frac{\delta\td}\tdq\right)
  -~\delta\cdt\,\left(\frac1\tu -\frac1\td\right)\,\right]~,\nn\\
\eea 
where $\delta\sdt = 2\,\cdt\,\delta \theta_t$ and $\delta\cdt =
-2\,\sdt\,\delta \theta_t$. If the parameters in the top/stop sector
are renormalized in the OS scheme, the shifts $\delta m_t$, $\delta
\ti$, $\delta \theta_t$ and $\delta A_t$ can be found in appendix~B of
ref.~\cite{Degrassi:2001yf}.

\section{Extension to the NMSSM}

In this appendix we describe how our results for the box form factor
for Higgs pair production in the MSSM can be extended to the case of
the NMSSM. Instead of the Higgs mass term $\mu\,H_1 H_2$, which in the
simplest realization of the NMSSM is forbidden by a $Z_3$ symmetry,
the superpotential contains\,\footnote{For consistency with the
  definition of $\mu$ in our MSSM results, here we adopt for the sign
  of $\lambda$ the opposite convention with respect to
  ref.~\cite{Degrassi:2009yq} and most public codes for NMSSM
  calculations. We also note that our normalization of the EW
  parameter, $v\approx 246$~GeV, differs by a factor $\sqrt{2}$ from the
  one in ref.~\cite{Degrassi:2009yq}.}
\be
\label{superpotential}
W ~\supset~ \lambda\,S H_1 H_2 ~+~ \frac{\kappa}{3}\,S^3~,
\ee
where $S$ is an additional gauge-singlet superfield. An effective
$\mu$ term is generated by the singlet VEV as $\mu = \lambda\,\langle
S\rangle$, and the CP-even parts $S_i$ of the neutral component of the
three Higgs fields -- ordered as $(H_1, H_2, S)$ -- mix into three
mass eigenstates which we denote as $h_a$,
\be
h_a ~=~ R^\smallS_{ai} \,S_i~,
\ee
where $R^\smallS$ is an orthogonal matrix. The decompositions of the
triangle and box form factors in eqs.~(\ref{Fdeltah})--(\ref{FBoxhH})
generalize to
\be
F_\Delta^{h_a} ~=\, - T_F\,
R^\smallS_{ai}\,{\cal H}_{i}~,
~~~~~~~~~~~~
F_\Box^{h_a h_b} ~=\, - T_F\,
R^\smallS_{ai}\,R^\smallS_{bj}\,{\cal H}_{ij}~.
\ee

The extension to the NMSSM of the results of
refs.~\cite{Degrassi:2008zj, Degrassi:2010eu, Degrassi:2012vt} for the
triangle form factors of the MSSM has been presented, in the context
of single Higgs production, in ref.~\cite{Liebler:2015bka}. Concerning
the box form factors, the terms ${\cal H}_{11}$, ${\cal H}_{12}$ and
${\cal H}_{22}$ coincide with those obtained for the MSSM in
section~\ref{sec:formfactors}. The top/stop contributions to the
remaining terms read
\bea
\label{H13}
{\cal H}^t_{13}& = &
\frac{\sq2\,\lambda\,v\,m_t}{\sin\beta}\,\left[
\frac12\,\mt\,\mu\,\cot\beta\,\sdt^2\,F_3 ~+~
\frac{\mt\,(A_t + 2 \,\mu\,\cot\beta)}{\tu-\td}\,F
~+~ \mz^2\,\cos^2\beta \,\sdt\,\widetilde F_2\,\right]\,,\\[3mm]
\label{H23}
{\cal H}^t_{23}& = &
\frac{\sq2\,\lambda\,v\,m_t}{\sin\beta}\,\left[\mt^2\,\cot\beta\,\sdt\,F_2
~+~\frac12\,\mt\,A_t\,\cot\beta\,\sdt^2\,F_3 ~+~
\frac{\mt\,A_t\,\cot\beta}{\tu-\td}\,F\right.\nn\\
&&\left.\phantom{\frac{\mt\,A_t\,\cot\beta}{\tu-\td}}~~
-\, \mz^2\,\sin\beta\cos\beta \,\sdt\,\widetilde F_2\,\right]\,,\\[3mm]
\label{H33}
{\cal H}^t_{33}& = &
~\lambda^2\,v^2~\left[
\frac12\,\mt^2\,\cot^2\beta\,\sdt^2\,F_3 ~+~
\frac{\mt^2\,\cot^2\beta}{\tu-\td}\,F\,\right]~,
\eea
where the functions $F_2$, $F_3$, $F$ and $\widetilde F_2$ coincide
with those entering the MSSM results, see
section~\ref{sec:formfactors} and appendix~\ref{app:functions}. In the
limit $\mb = \theta_b = 0$ there are no contributions to
${\cal H}_{13}$, ${\cal H}_{23}$ and ${\cal H}_{33}$ from
bottom/sbottom loops.

Finally, when the parameters entering the top/stop contributions to
the one-loop part of the form factors are expressed in a
renormalization scheme other than $\drbar$, the shifts to the form
factors that were not already given in section~\ref{OSshift} read
\bea
\label{dH13}
\delta{\cal H}^t_{13}& = &
\frac{\sq2\,\lambda\,v\,m_t}{\sin\beta}\,\left[
\frac12\,\mt\,\mu\,\cot\beta\,\sdt^2\,\delta F_3 ~+~
\frac{\mt\,(A_t + 2 \,\mu\,\cot\beta)}{\tu-\td}\,\delta F
~+~ \mz^2\,\cos^2\beta \,\sdt\, \delta \widetilde F_2\,\right.\nn\\[1mm]
&&~~~~~~~~~~~~~~~~
+\left. \frac{\mt\,\delta A_t}{\tu-\td}\,F^{1\ell}\,\right]\,,\\[3mm]
\label{dH23}
\delta{\cal H}^t_{23}& = &
\frac{\sq2\,\lambda\,v\,m_t}{\sin\beta}\,\left[
\mt^2\,\cot\beta\,\sdt\,\delta F_2
~+~\frac12\,\mt\,A_t\,\cot\beta\,\sdt^2\,\delta F_3 ~+~
\frac{\mt\,A_t\,\cot\beta}{\tu-\td}\,\delta F\,\right.\nn\\[1mm]
&&~~~~~~~~~~~~~~~
\left. -\, \mz^2\,\sin\beta\cos\beta \,\sdt\,\delta \widetilde F_2
~+~\frac12\,\mt\,\delta A_t\,\cot\beta\,\sdt^2\,F_3^{1\ell}
~+~\frac{\mt\,\delta A_t\,\cot\beta}{\tu-\td}\, F^{1\ell}\,\right]\,,\nn\\\\
\label{dH33}
\delta {\cal H}^t_{33}& = &
~\lambda^2\,v^2\,~\left[
\frac12\,\mt^2\,\cot^2\beta\,\sdt^2\,\delta F_3 ~+~
\frac{\mt^2\,\cot^2\beta}{\tu-\td}\,\delta F\,\right]~,
\eea
where the shifts $\delta F_2$, $\delta F_3$, $\delta F$ and $\delta
\widetilde F_2$ coincide with those defined in
appendix~\ref{app:shifts}.

\end{appendices}

\vfill
\newpage 

\bibliographystyle{utphys}
\bibliography{ADGS}

\end{document}

%% file: boxformfac
\begingroup
  \makeatletter
  \providecommand\color[2][]{%
    \GenericError{(gnuplot) \space\space\space\@spaces}{%
      Package color not loaded in conjunction with
      terminal option `colourtext'%
    }{See the gnuplot documentation for explanation.%
    }{Either use 'blacktext' in gnuplot or load the package
      color.sty in LaTeX.}%
    \renewcommand\color[2][]{}%
  }%
  \providecommand\includegraphics[2][]{%
    \GenericError{(gnuplot) \space\space\space\@spaces}{%
      Package graphicx or graphics not loaded%
    }{See the gnuplot documentation for explanation.%
    }{The gnuplot epslatex terminal needs graphicx.sty or graphics.sty.}%
    \renewcommand\includegraphics[2][]{}%
  }%
  \providecommand\rotatebox[2]{#2}%
  \@ifundefined{ifGPcolor}{%
    \newif\ifGPcolor
    \GPcolortrue
  }{}%
  \@ifundefined{ifGPblacktext}{%
    \newif\ifGPblacktext
    \GPblacktexttrue
  }{}%
  \let\gplgaddtomacro\g@addto@macro
  \gdef\gplbacktext{}%
  \gdef\gplfronttext{}%
  \makeatother
  \ifGPblacktext
    \def\colorrgb#1{}%
    \def\colorgray#1{}%
  \else
    \ifGPcolor
      \def\colorrgb#1{\color[rgb]{#1}}%
      \def\colorgray#1{\color[gray]{#1}}%
      \expandafter\def\csname LTw\endcsname{\color{white}}%
      \expandafter\def\csname LTb\endcsname{\color{black}}%
      \expandafter\def\csname LTa\endcsname{\color{black}}%
      \expandafter\def\csname LT0\endcsname{\color[rgb]{1,0,0}}%
      \expandafter\def\csname LT1\endcsname{\color[rgb]{0,1,0}}%
      \expandafter\def\csname LT2\endcsname{\color[rgb]{0,0,1}}%
      \expandafter\def\csname LT3\endcsname{\color[rgb]{1,0,1}}%
      \expandafter\def\csname LT4\endcsname{\color[rgb]{0,1,1}}%
      \expandafter\def\csname LT5\endcsname{\color[rgb]{1,1,0}}%
      \expandafter\def\csname LT6\endcsname{\color[rgb]{0,0,0}}%
      \expandafter\def\csname LT7\endcsname{\color[rgb]{1,0.3,0}}%
      \expandafter\def\csname LT8\endcsname{\color[rgb]{0.5,0.5,0.5}}%
    \else
      \def\colorrgb#1{\color{black}}%
      \def\colorgray#1{\color[gray]{#1}}%
      \expandafter\def\csname LTw\endcsname{\color{white}}%
      \expandafter\def\csname LTb\endcsname{\color{black}}%
      \expandafter\def\csname LTa\endcsname{\color{black}}%
      \expandafter\def\csname LT0\endcsname{\color{black}}%
      \expandafter\def\csname LT1\endcsname{\color{black}}%
      \expandafter\def\csname LT2\endcsname{\color{black}}%
      \expandafter\def\csname LT3\endcsname{\color{black}}%
      \expandafter\def\csname LT4\endcsname{\color{black}}%
      \expandafter\def\csname LT5\endcsname{\color{black}}%
      \expandafter\def\csname LT6\endcsname{\color{black}}%
      \expandafter\def\csname LT7\endcsname{\color{black}}%
      \expandafter\def\csname LT8\endcsname{\color{black}}%
    \fi
  \fi
  \setlength{\unitlength}{0.0500bp}%
  \begin{picture}(7200.00,5040.00)%
    \gplgaddtomacro\gplbacktext{%
      \csname LTb\endcsname%
      \put(946,704){\makebox(0,0)[r]{\strut{}-4}}%
      \put(946,1213){\makebox(0,0)[r]{\strut{}-3.5}}%
      \put(946,1722){\makebox(0,0)[r]{\strut{}-3}}%
      \put(946,2231){\makebox(0,0)[r]{\strut{}-2.5}}%
      \put(946,2740){\makebox(0,0)[r]{\strut{}-2}}%
      \put(946,3248){\makebox(0,0)[r]{\strut{}-1.5}}%
      \put(946,3757){\makebox(0,0)[r]{\strut{}-1}}%
      \put(946,4266){\makebox(0,0)[r]{\strut{}-0.5}}%
      \put(946,4775){\makebox(0,0)[r]{\strut{} 0}}%
      \put(1078,484){\makebox(0,0){\strut{} 500}}%
      \put(1651,484){\makebox(0,0){\strut{} 600}}%
      \put(2223,484){\makebox(0,0){\strut{} 700}}%
      \put(2796,484){\makebox(0,0){\strut{} 800}}%
      \put(3368,484){\makebox(0,0){\strut{} 900}}%
      \put(3941,484){\makebox(0,0){\strut{} 1000}}%
      \put(4513,484){\makebox(0,0){\strut{} 1100}}%
      \put(5086,484){\makebox(0,0){\strut{} 1200}}%
      \put(5658,484){\makebox(0,0){\strut{} 1300}}%
      \put(6231,484){\makebox(0,0){\strut{} 1400}}%
      \put(6803,484){\makebox(0,0){\strut{} 1500}}%
      \put(176,2739){\rotatebox{-270}{\makebox(0,0){\strut{}$F_{\Box}^{hh}$}}}%
      \put(3940,154){\makebox(0,0){\strut{}$M_{S}$ [GeV]}}%
    }%
    \gplgaddtomacro\gplfronttext{%
      \csname LTb\endcsname%
      \put(5662,1996){\makebox(0,0)[r]{\strut{}1-loop, top only}}%
      \csname LTb\endcsname%
      \put(5662,1688){\makebox(0,0)[r]{\strut{}1-loop, top+SUSY}}%
      \csname LTb\endcsname%
      \put(5662,1380){\makebox(0,0)[r]{\strut{}2-loop, top only}}%
      \csname LTb\endcsname%
      \put(5662,1072){\makebox(0,0)[r]{\strut{}2-loop, top+SUSY}}%
    }%
    \gplbacktext
    \put(0,0){\includegraphics{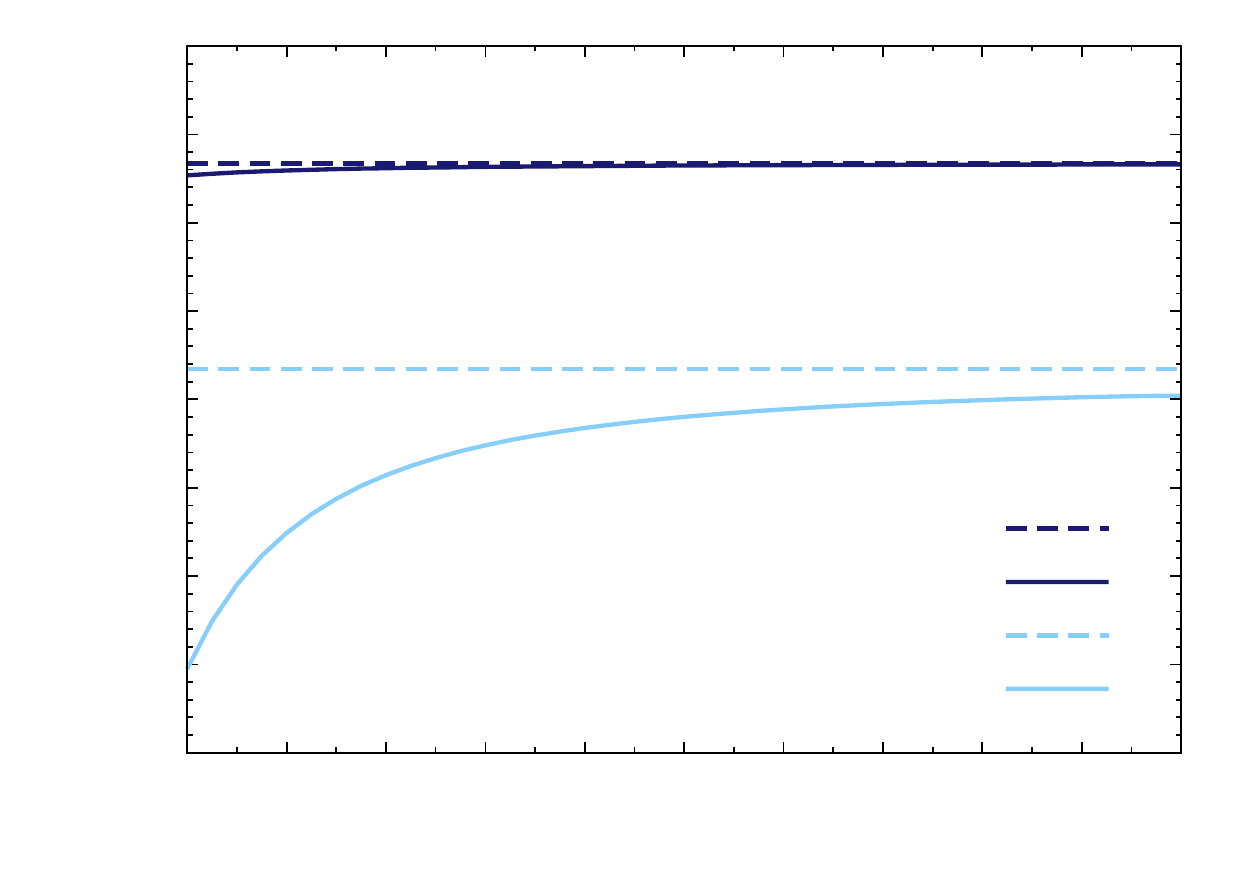}}%
    \gplfronttext
  \end{picture}%
\endgroup

%% file: cxn
\begingroup
  \makeatletter
  \providecommand\color[2][]{%
    \GenericError{(gnuplot) \space\space\space\@spaces}{%
      Package color not loaded in conjunction with
      terminal option `colourtext'%
    }{See the gnuplot documentation for explanation.%
    }{Either use 'blacktext' in gnuplot or load the package
      color.sty in LaTeX.}%
    \renewcommand\color[2][]{}%
  }%
  \providecommand\includegraphics[2][]{%
    \GenericError{(gnuplot) \space\space\space\@spaces}{%
      Package graphicx or graphics not loaded%
    }{See the gnuplot documentation for explanation.%
    }{The gnuplot epslatex terminal needs graphicx.sty or graphics.sty.}%
    \renewcommand\includegraphics[2][]{}%
  }%
  \providecommand\rotatebox[2]{#2}%
  \@ifundefined{ifGPcolor}{%
    \newif\ifGPcolor
    \GPcolortrue
  }{}%
  \@ifundefined{ifGPblacktext}{%
    \newif\ifGPblacktext
    \GPblacktexttrue
  }{}%
  \let\gplgaddtomacro\g@addto@macro
  \gdef\gplbacktext{}%
  \gdef\gplfronttext{}%
  \makeatother
  \ifGPblacktext
    \def\colorrgb#1{}%
    \def\colorgray#1{}%
  \else
    \ifGPcolor
      \def\colorrgb#1{\color[rgb]{#1}}%
      \def\colorgray#1{\color[gray]{#1}}%
      \expandafter\def\csname LTw\endcsname{\color{white}}%
      \expandafter\def\csname LTb\endcsname{\color{black}}%
      \expandafter\def\csname LTa\endcsname{\color{black}}%
      \expandafter\def\csname LT0\endcsname{\color[rgb]{1,0,0}}%
      \expandafter\def\csname LT1\endcsname{\color[rgb]{0,1,0}}%
      \expandafter\def\csname LT2\endcsname{\color[rgb]{0,0,1}}%
      \expandafter\def\csname LT3\endcsname{\color[rgb]{1,0,1}}%
      \expandafter\def\csname LT4\endcsname{\color[rgb]{0,1,1}}%
      \expandafter\def\csname LT5\endcsname{\color[rgb]{1,1,0}}%
      \expandafter\def\csname LT6\endcsname{\color[rgb]{0,0,0}}%
      \expandafter\def\csname LT7\endcsname{\color[rgb]{1,0.3,0}}%
      \expandafter\def\csname LT8\endcsname{\color[rgb]{0.5,0.5,0.5}}%
    \else
      \def\colorrgb#1{\color{black}}%
      \def\colorgray#1{\color[gray]{#1}}%
      \expandafter\def\csname LTw\endcsname{\color{white}}%
      \expandafter\def\csname LTb\endcsname{\color{black}}%
      \expandafter\def\csname LTa\endcsname{\color{black}}%
      \expandafter\def\csname LT0\endcsname{\color{black}}%
      \expandafter\def\csname LT1\endcsname{\color{black}}%
      \expandafter\def\csname LT2\endcsname{\color{black}}%
      \expandafter\def\csname LT3\endcsname{\color{black}}%
      \expandafter\def\csname LT4\endcsname{\color{black}}%
      \expandafter\def\csname LT5\endcsname{\color{black}}%
      \expandafter\def\csname LT6\endcsname{\color{black}}%
      \expandafter\def\csname LT7\endcsname{\color{black}}%
      \expandafter\def\csname LT8\endcsname{\color{black}}%
    \fi
  \fi
  \setlength{\unitlength}{0.0500bp}%
  \begin{picture}(7200.00,5040.00)%
    \gplgaddtomacro\gplbacktext{%
      \csname LTb\endcsname%
      \put(814,704){\makebox(0,0)[r]{\strut{} 10}}%
      \put(814,1518){\makebox(0,0)[r]{\strut{} 20}}%
      \put(814,2332){\makebox(0,0)[r]{\strut{} 30}}%
      \put(814,3147){\makebox(0,0)[r]{\strut{} 40}}%
      \put(814,3961){\makebox(0,0)[r]{\strut{} 50}}%
      \put(814,4775){\makebox(0,0)[r]{\strut{} 60}}%
      \put(946,484){\makebox(0,0){\strut{} 500}}%
      \put(1532,484){\makebox(0,0){\strut{} 600}}%
      \put(2117,484){\makebox(0,0){\strut{} 700}}%
      \put(2703,484){\makebox(0,0){\strut{} 800}}%
      \put(3289,484){\makebox(0,0){\strut{} 900}}%
      \put(3875,484){\makebox(0,0){\strut{} 1000}}%
      \put(4460,484){\makebox(0,0){\strut{} 1100}}%
      \put(5046,484){\makebox(0,0){\strut{} 1200}}%
      \put(5632,484){\makebox(0,0){\strut{} 1300}}%
      \put(6217,484){\makebox(0,0){\strut{} 1400}}%
      \put(6803,484){\makebox(0,0){\strut{} 1500}}%
      \put(176,2739){\rotatebox{-270}{\makebox(0,0){\strut{}$\sigma (pp\to hh)$ [fb]}}}%
      \put(3874,154){\makebox(0,0){\strut{}$M_{S}$ [GeV]}}%
    }%
    \gplgaddtomacro\gplfronttext{%
      \csname LTb\endcsname%
      \put(5655,4458){\makebox(0,0)[r]{\strut{}LO, full}}%
      \csname LTb\endcsname%
      \put(5655,4150){\makebox(0,0)[r]{\strut{}LO, no SUSY}}%
      \csname LTb\endcsname%
      \put(5655,3842){\makebox(0,0)[r]{\strut{}NLO, full}}%
      \csname LTb\endcsname%
      \put(5655,3534){\makebox(0,0)[r]{\strut{}NLO, no SUSY}}%
      \csname LTb\endcsname%
      \put(5655,3226){\makebox(0,0)[r]{\strut{}NLO, no SUSY in $F^{2\ell}_{\Box}$}}%
    }%
    \gplbacktext
    \put(0,0){\includegraphics{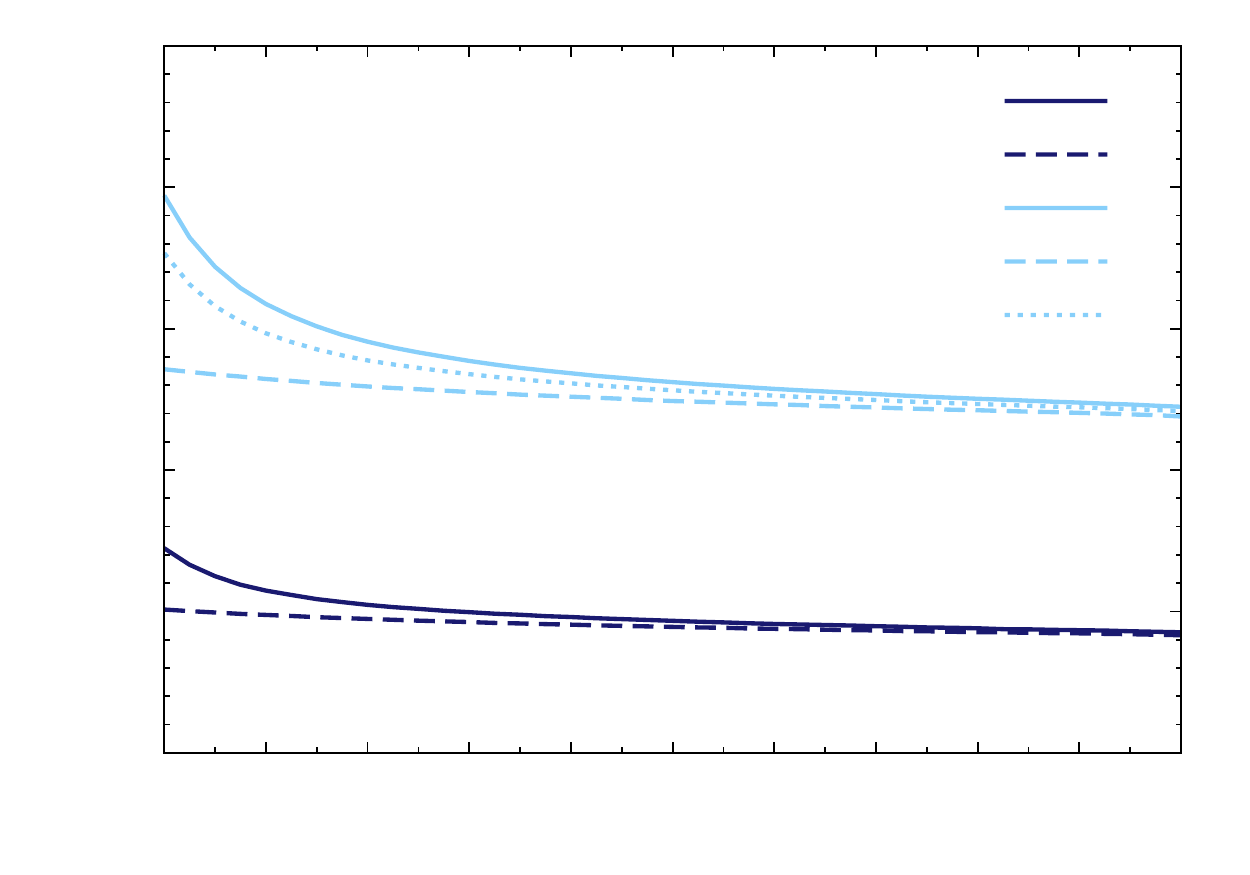}}%
    \gplfronttext
  \end{picture}%
\endgroup